\documentclass[prd,twocolumn,nofootinbib,showpacs]{revtex4}
\usepackage{graphicx}
\usepackage{epsfig}
\usepackage{color}
\usepackage{mathrsfs}
\usepackage{amsmath}
\usepackage{amssymb}

\newcommand{\be}{\begin{equation}}
\newcommand{\ee}{\end{equation}}
\newcommand{\bd}{\begin{equation*}}
\newcommand{\ed}{\end{equation*}}
\newcommand{\bea}{\begin{eqnarray}}
\newcommand{\eea}{\end{eqnarray}}
\newcommand{\ra}{\rightarrow}
\newcommand{\hs}{\hspace{2mm}}

\newcommand{\gapp}{\mathrel{\raise.3ex\hbox{$>$}\mkern-14mu
              \lower0.6ex\hbox{$\sim$}}}
\newcommand{\lapp}{\mathrel{\raise.3ex\hbox{$<$}\mkern-14mu
              \lower0.6ex\hbox{$\sim$}}}

\begin{document}

\title{Quantum kinetics and prethermalization of Hawking radiation}
\author{Dmitry Podolsky}
\email{podolsky@phys.cwru.edu}
\author{Eric Greenwood}
\author{Glenn Starkman}
\affiliation{Department of Physics \& CERCA, Case Western Reserve University, Cleveland, OH 44106-7079}
\begin{abstract}

 We reinvestigate the emission of Hawking radiation during  gravitational collapse to a black hole. Both CGHS collapse of a shock wave in $(1+1)$-dimensional dilaton gravity and Schwarzschild collapse of a spherically symmetric thin shell in $(3+1)$-dimensional gravity are considered.  Studying the dynamics of in-vacuum polarization, we find that a multi-parametric family of out-vacua exists. Initial conditions for the collapse lead dynamically to different vacua from this family as the final state.   Therefore, the form of the out-vacuum  encodes memory about the  initial quantum state of the system. While most out-vacua feature a non-thermal Hawking flux and are expected to decay quickly, there also exists a thermal vacuum state. Collectively, these observations suggest an interesting possible resolution of the information loss paradox.

\end{abstract}

 \maketitle

\section{Introduction}

Hawking's discovery of black hole radiance \cite{Hawking:1974sw} in 1974, along with Bekenstein's conjecture \cite{Bekenstein:1973ur} of the generalized second law of thermodynamics in 1973 inspired a conceptually pleasing connection between general relativity, thermodynamics and quantum theory.  It also led to a serious conundrum. As Hawking showed, when a quantum field theory is put on a fixed black hole background, an observer at spatial infinity will detect a flux of the quanta of that field in a mixed quantum state described by a thermal density matrix. This thermal flux leads in time to the complete evaporation of the black hole. Thus, if matter quanta are thrown into an already-formed black hole, any quantum information encoded in their wave function is lost to the outside once those quanta cross the horizon. Because the Hawking radiation itself encodes no information other than the size of the black hole, the in-fallen information cannot be recovered as the black hole evaporates. Thus,  once the black hole evaporated completely, a pure incoming quantum state has been transformed into a mixed outgoing quantum state in apparent contradiction with the unitarity of quantum mechanics. This is the essence of the famous information loss paradox \cite{Hawking:1976ra,Preskill:1992tc}.

The black-hole information loss  paradox is one of the most pressing problems at the interface between quantum mechanics and General Relativity, and it does not seem that one can appeal to quantum gravity for a solution. The quanta of the Hawking radiation are produced in the vicinity of horizon, where the gravitational field is weak for macroscopic black holes; therefore, the approximation of doing quantum field theory on a fixed curved background spacetime should be valid. Neither can one expect the paradox to somehow be resolved once the evaporating black hole becomes of Planckian size and quantum gravity does kick in. There is far too much information encoded by the in-falling matter of a stellar black hole to be stored in a Planckian remnant -- the information must be released long before quantum gravity should matter.

In an attempt to understand the information loss problem in the context of quantum gravity,  Callan, Giddings, Harvey and Strominger (CGHS) studied $(1+1)$-dimensional dilatonic black holes \cite{Callan:1992rs}. The major advantages to the CGHS model are that, unlike the 4-dimensional Einstein gravity case, (a)  $(1+1)$-dimensional quantum gravity is a renormalizable theory and (b) the expectation value of the energy-momentum tensor of the Hawking radiation is directly related to the conformal anomaly of the theory \cite{Christensen:1977jc} and is therefore explicitly calculable. This allows back-reaction to be included into the model relatively easily. Once this is done, information still appears to be lost \cite{Fiola:1994ir} due to the emission of thermal Hawking radiation.\footnote{Recently, evaporation of 2-dimensional CGHS black holes with backreaction effects taken into account has been revisited in \cite{Ashtekar:2010qz}, revealing a non-thermal power-law-like spectrum of Hawking radiation.}

There have been several classes of proposed resolutions of the information loss problem. Most (if not all) of these proposals  \cite{Mathur:2009hf} seem to bring with them other physical or conceptual difficulties:
 \newline
(1) Information really is lost \cite{Hawking:1976ra}, implying that our present laws of quantum
mechanics are insufficient to address the formation and evaporation of a black hole. Thus one
either has to abandon the idea of energy conservation by introducing a non-unitary time
evolution or allow quantum information to dissipate and therefore ultimately violate unitarity of quantum mechanics (see for example \cite{'tHooft:1999gk}). \newline
(2) Information is not lost; instead, it is stored inside the black hole.  The black hole does not evaporate completely, but leaves behind a remnant that has a mass on the order of $M_P$ \cite{Aharonov:1987tp}. This remnant contains all the original information. There are two major problems with this idea: (a) it requires infinitely many species of remnants (since there are infinitely many initial states leading to gravitational collapse), which, in addition to aesthetic objections, leads to an unacceptably high rate of pair production of such remnants in a gravitational field \cite{Giddings:1994pj}; and (b) there seems to be no mechanism to stabilize such remnants. A related idea  is that a ``baby universe'' is formed within the black hole and the in-falling information gets transferred to that universe. \newline
(3) The physics of a black hole and of Hawking radiation is  non-local (see for example \cite{Giddings:2009ae,Giddings:2012bm}). Deviations from locality would automatically imply corrections to the Hawking result. One way to introduce this non-locality is to assume that the degrees of freedom inside a black hole are duplicated in the exterior \cite{Balasubramanian:2011dm,STU}. The information thereby remains intact in the exterior. This however seems to be at odds with the superposition principle of quantum mechanics \cite{Strominger:1994tn}. Still, the idea of non-local physics somehow resolving the information-loss paradox presently seems to be the most popular in the literature. \newline
(4) Information escapes during collapse and evaporation. This can happen if information is subtly encoded into correlations among the particles that fall into and escape the black hole or if, contrary to Hawking's calculation, the Hawking radiation is not completely thermal \cite{Israel:2010gk} and can carry out small amounts of information. It is possible that (4) is automatically included in (3), once we learn how to describe the non-local physics of gravitational collapse.

In this paper we try to address both possibilities (3) and (4) for the scenarios of CGHS collapse as well as $(3+1)$-dimensional spherically symmetric Schwarzschild collapse according to the following program:
\begin{itemize}
\item All the calculations will be done in the Schwarzschild slicing, where the collapse takes an infinite amount of time to complete but the collapsing shell of matter approaches its apparent horizon exponentially quickly.
\item Following Hawking, we shall quantize the scalar field $f(t,r)$ on the fixed background spacetime appropriate to the collapse, neglecting back-reaction effects. The difference from the standard approach is that we shall use Heisenberg quantization, where the modes of the field are considered time-independent while their time-dependent amplitudes are promoted to quantum-mechanical operators.
\item Using the results of Heisenberg quantization, we shall study \emph{the dynamics} of in-vacuum polarization in order to find how and when the out-vacuum of the scalar field is formed. We shall also determine the explicit form of the out-vacuum.
\item Finally, we shall calculate the Bogolyubov coefficients between the in-vacuum and the formed out-vacuum in order to find the spectrum of Hawking radiation. We shall also calculate the response function of a detector carried by a fiducial observer located at spatial infinity.
\end{itemize}

As we shall see, this program leads to several surprising results. We find that an out-vacuum is formed as a Bose-Einstein condensate of ``in'' excitations within a time interval $\delta{}t\sim{}R_S$ after the onset of the collapse. Instead of a unique out-vacuum, there exists a continuous family of ``out'' vacua. Which particular vacuum from this family becomes the post-collapse final state depends on the initial state of the collapse process. A calculation of the response function of a detector shows that, for most of these vacua, the Hawking flux is not thermal (and divergent), while for one particular vacuum from the family it is thermal (and finite). This fact has important implications for the information loss paradox.

The paper is organized as follows. In Section \ref{sec:gravcollapse} we briefly review the classical gravitational collapse of a thin shell of matter in both $(1+1)$-dimensional dilaton gravity (Section \ref{CGHSclassical}) and $(3+1)$-dimensional general relativity (Section \ref{4Dclassical}). 
We perform Heisenberg quantization of the scalar field on a fixed collapsing CGHS background in Section \ref{sec:CGHSmodes}. 
To check that the quantization was performed correctly, we assume a particular form for the out-vacuum and calculate Bogolyubov coefficients between in- and out-vacua in Section \ref{sec:CGHSBogolyubov}. The calculated Hawking flux reproduces the classical result by Giddings and Nelson \cite{Giddings:1992ff}. Then, in Section \ref{sec:CGHSQuantumKineticOUT} we derive the quantum kinetic equation describing Bose-Einstein condensation of ``in'' excitations and the process of formation of the out-vacuum. We find the explicit form of the out-vacuum. The same analysis is repeated for $(3+1)$-dimensional spherically symmetric Schwarzschild collapse in Section \ref{sec:MSF}. 
We summarize our results and muse extensively on the resolution of the information loss paradox in Section \ref{Conclusion}. In particular, subsection \ref{sec:mainpoints} contains the review of our major results. Physical content of quantum kinetic equations discussing particle production in QFTs on fixed curved background is discussed in subsection \ref{sec:QKnotes}. Properties of out-vacua and physical observables are discussed in subsection \ref{sec:VacNotes} and implications of our results for information loss paradox --- in subsection \ref{sec:IL}. Finally, the collapse of a thin spherically symmetric shell on the formed primordial black hole is considered in Appendix \ref{sec:PBH}, while the case of two concentric collapsing shells is discussed in Appendix \ref{sec:2Shells}. In Appendix \ref{sec:WKBnotes} we explain how our results help to resolve the famous issue of WKB breakdown in the near horizon regime for solutions of the Wheeler-de Witt equation describing collapse and Hawking evaporation.

\section{Gravitational collapse in classical theory}
\label{sec:gravcollapse}

We first review some of the aspects of the classical solution describing a thin shell collapsing to a  black hole in two-dimensional dilaton gravity \cite{Callan:1992rs} and four-dimensional Einstein gravity (with spherical symmetry). Our major concern will be to compute the dynamics of collapse in Schwarzschild-like coordinate systems where the apparent horizon, as seen by a fiducial observer at spatial infinity, is spacelike, full collapse time (the time before the collapsing body/shell reaches its event horizon) is infinite, but the approach of the shell to its apparent horizon is exponentially rapid.

\subsection{CGHS collapse: classical solution}
\label{CGHSclassical}

Two-dimensional dilaton gravity is the theory described by the action
\be
  S\!=\!\int\!\frac{d^2x}{2\pi}\sqrt{-g}\left[e^{-\phi}(R+4(\nabla\phi)^2+4\lambda^2)-(\nabla f)^2\right],
  \label{Sdilaton}
\ee
where $g$, $\phi$, and $f$ are the metric, dilaton, and matter fields, respectively, while
$\lambda^2$ is a cosmological constant (see \cite{Grumiller:2002nm} for comprehensive review).
The equation of motion for the dilaton $\phi$ is given by
\be
  0=2e^{-2\phi}\left(\nabla_{\mu}\nabla_{\nu}\phi+g_{\mu\nu}\left(\nabla^2\phi+\lambda^2-(\nabla\phi)^2\right)\right),
  \label{EOMphi}
\ee
while the dynamics of spacetime is determined by
\be
  0=2e^{-2\phi}\left(4\nabla^2\phi+4\lambda^2+R-4(\nabla\phi)^2\right).
  \label{EOMg}
\ee

It is convenient to analyze the theory in conformal gauge and introduce light-cone coordinates $x^{\pm}=x^0\pm x^1$. Then, the metric takes the form
\be
  ds^2=-e^{2\rho}dx^+dx^-
  \label{rho_met}
\ee
with metric coefficients
\be
  g_{+-}=g_{-+}=-\frac{1}{2}e^{2\rho}.
  \label{g_comps}
\ee
The Ricci scalar $R$ present in (\ref{EOMg}) is given by
\be
  R=8e^{-2\rho}\partial_+\partial_-\rho.
  \label{Ricci}
\ee

In terms of $\rho$, $\phi$ and $f$, the equations of motion 
can then be written as
\begin{align}
  \hs 0=&e^{-2(\phi+\rho)}\left(\lambda^2e^{2\rho}-4\partial_+\partial_-\phi+2\partial_+\partial_-\rho+\right.\nonumber\\
  &+\left.4\partial_+\phi\partial_-\phi\right)\label{EOMPHI}\\
  \hs 0=&e^{-2\phi}\left(2\partial_+\partial_-\phi-4\partial_+\phi\partial_-\phi-\lambda^2e^{2\rho}\right)\label{EOMrho}\\
  \hs 0=&\partial_+\partial_-f\label{EOMf},
\end{align}

The $g_{++}$ and $g_{--}$ components give constraint equations
\be
  e^{-2\phi}\left(4\partial_{\pm}\rho\partial_{\pm}\phi-2\partial_{\pm}^2\phi\right)=-\frac{1}{2}(\partial_{\pm}f)^2.
  \label{constraint}
\ee

Since the gauge is not fixed completely by the condition (\ref{g_comps}), we can choose the Kruskal condition
\be
  \phi=\rho,
\ee
so that the metric (\ref{rho_met}) reduces to
\be
  ds^2=-e^{2\phi}dx^+dx^-.
\ee
The Eq. (\ref{EOMrho}) now acquires the form
\be
  \partial_+\partial_-(e^{-2\phi})=-\lambda^2
\ee
and upon solving yields
\be
  e^{-2\rho}=\frac{M}{\lambda}-\lambda^2x^+x^-,
\ee
where $M/\lambda$ is an integration constant. (The linear terms in $x^\pm$ are omitted, since they represent only constant shifts
in those coordinates.) $M$ is interpreted to be the mass of an eternal black hole.
In the limit $M=0$, this solution is known as the linear dilaton vacuum.

If one starts with the linear dilaton vacuum and then sends a pulse of the matter scalar field $f$,
a black hole is produced. For a sharp left-moving pulse, we can write
\be
  T_{++}^f=\frac{1}{2}(\partial_+f)^2=\frac{M}{\lambda x_0^+}\delta(x^+-x_0^+).
\ee
From (\ref{constraint}), again choosing $\rho=\phi$, the general solution for $e^{-2\phi}$
is given by
\begin{align}
  e^{-2\phi}&=-\lambda^2x^+x^--\frac{M}{\lambda x_0^+}\int dx'^+\int^{x'^+}dx''^+\delta(x''^+-x_0^+)\nonumber\\
                   &=-\lambda^2x^+x^--\frac{M}{\lambda x_0^+}(x^+-x_0^+)\Theta(x^+-x_0^+),
\end{align}
where $\Theta(x-y)$ is the Heaviside theta function. Thus for $x^+<x_0^+$ this solution
corresponds to the linear dilaton vacuum solution, while for $x^+>x_0^+$ it describes the
black hole of the mass $M$. 
The spacetime metric takes the form
\be
  ds^2=-\frac{dx^+dx^-}{-\lambda^2x^+x^--\frac{M}{\lambda x_0^+}(x^+-x_0^+)\Theta(x^+-x_0^+)}
\label{eq:CGHScollapsemetric}
\ee

\subsection{CGHS collapse: coordinate systems}
\label{CGHScoord}

Here we will discuss the solution describing the $(1+1)$-dimensional collapsing shell in three different coordinate systems: Kruskal, Eddington-Finkelstein
and Schwarzschild \cite{SusskindLindesay}.

According to the expression (\ref{eq:CGHScollapsemetric}), the solution has different forms in the regions $x^+<x_0^+$ and $x^+>x_0^+$.

\subsubsection{$x^+<x_0^+$ region}

Inside the collapsing shell one has $e^{-2\phi}=-\lambda^2x^+x^-$, so that the metric is
\be
  ds^2=-\frac{dx^+dx^-}{\lambda^2x^+x^-}.
\ee
This coordinate system is a direct analogue of the Kruskal-Szekeres coordinate system for the $(3+1)$-D collapsing spherically symmetric shell. For a more physical interpretation, one can go to the Eddington-Finkelstein coordinates, which are defined by $x^{\pm}=\pm e^{\pm\lambda\sigma^{\pm}}$. The metric then is given by
\be
  ds^2=-d\sigma^+d\sigma^-,
\ee
so that the spacetime inside the collapsing shell is just flat Minkowski spacetime. For the Eddington-Finkelstein coordinate system, the inside region corresponds to
$\lambda\sigma^+<\ln(\lambda x_0^+)$.

\begin{figure}[htp]
\includegraphics[width=0.5\textwidth,height=0.3\textheight]{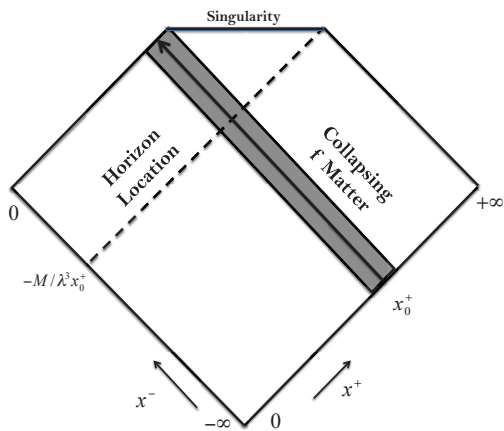}
\caption{Penrose diagram describing the CGHS collapsing setup.}
\label{PenroseCGHS}
\end{figure}

Writing $\sigma^+=t_{\sigma}+r_{\sigma}$, we find that the region $x^+<x_0^+$ corresponds to $r_{\sigma}<\lambda^{-1}\ln(\lambda x_0^+)-t_{\sigma}$.
The equation $r_{\sigma}=\lambda^{-1}\ln(\lambda x_0^+)-t_{\sigma}$ gives the position of the
moving shell from the point of view of an observer inside the shell. According to such an observer, the shell moves at the speed of light.

\subsubsection{$x^+>x_0^+$ region}

In the region outside the shell one has $e^{-2\phi}=M/\lambda-\lambda^2x^+x^-$ in the Kruskal-Szekeres coordinate system, so that the metric is
\be
  ds^2=-\frac{dx^+dx^-}{M/\lambda-\lambda^2x^+x^-}.
\ee
The Eddington-Finkelstein coordinates are now defined by
$x^+=e^{\lambda\sigma^+}$ and $e^{-\lambda\sigma^-}=-\lambda(x^-+M/\lambda^3)$, and
the metric is given as
\be
  ds^2=-\frac{d\sigma^+d\sigma^-}{1+(M/\lambda)e^{\lambda(\sigma^+-\sigma^-)}}.
\ee
The region $x^+>x_0^+$ in the Kruskal-Szekeres coordinates corresponds to $\lambda\sigma^+>\ln(\lambda x_0^+)$ or $r_{\sigma}>\lambda^{-1}\ln(\lambda x_0^+)-t_{\sigma}$ in the Eddington-Finkelstein coordinates.

The scalar curvature is given by
\be
R = \frac{4M\lambda}{M/\lambda - \lambda^2 x^+{}x^-} = \frac{4M\lambda}{M/\lambda+e^{2\lambda{}r_\sigma}},
\ee
where $r_\sigma = (\sigma^+-\sigma^-)/2$.

As we see, a singularity is reached at $x^+{}x^-=t^2-x^2=\frac{M}{\lambda^3}$ in the Kruskal-Szekeres coordinates or at $r_\sigma = \frac{1}{2\lambda}\log (-M/\lambda)$ in the Eddington-Finkelstein coordinates. One would naturally expect this singularity to be hidden by the event horizon and this is indeed what happens in this setup \cite{Strominger:1994tn}. If we assume for a moment that we are dealing with a $(3+1)$-dimensional spherically symmetric setup instead of $(1+1)$-dimensional one, the area of 2-spheres (corresponding to points in the $(1+1)$-dimensional setup) would be given by $\frac{4\pi}{\lambda^2}e^{-2\phi{}(x^+,x^-)}$. Based on the form of the solution for the dilaton, it is easy to check that all the 2-spheres (points) such that $x^-<-M/\lambda^3x_0^+\equiv x_H$, $x^+<x_0^+$ are trapped. The apparent horizon is the boundary of such region given by $x^-=x_H$. For the given setup, it coincides with the event horizon, so we shall denote it further simply as the horizon.

In the Eddington-Finkelstein coordinates, the horizon is located at $\sigma^+=-\infty$ (or $\sigma^-=+\infty$). From above, it is only reached as
\be
  e^{-\lambda(2t_{\sigma}-\lambda^{-1}\ln(\lambda x^+_0))}=\lambda x_0^+e^{-2\lambda t_{\sigma}}=0,
\ee
that is, at $t_{\sigma}\ra\infty$.

As we see, the BH formation takes infinite time in the Eddington-Finkelstein coordinates, but the approach to it is exponentially rapid. Since we would like to make parallels with a spherically symmetric $(3+1)$-dimensional gravitational collapse, this class of coordinate systems is naturally the only one that interests us.

Another important coordinate system where the dynamics of collapse has similar features is the Schwarzschild-like one. Let us define
\be
  r=\frac{1}{2\lambda}\ln\left(\frac{M}{\lambda}+e^{2\lambda r_{\sigma}}\right)+\frac{1}{2\lambda}\ln\left(\frac{e\lambda}{M}\right),
  \label{R_dil}
\ee
so that the metric can be rewritten in the form:
\be
  ds^2=-\left(1-e^{1-2\lambda r}\right)dt_{\sigma}^2+\frac{dr^2}{1-e^{1-2\lambda r}}.
  \label{DSM}
\ee
As usual, there is a coordinate singularity at $r=(2\lambda)^{-1}\equiv R_S$, corresponding to the location of the horizon.

Let us determine the equation for the motion of the shell in this coordinate system. Recall that the equation
$r_\sigma = \frac{1}{\lambda}\log{}(\lambda{}x^+_0)-t_\sigma$ describes the motion of the shell in Eddington-Finkelstein coordinates; hence in Schwarzchild-like coordinates, we find that
\begin{align}
  R(t)&=R_S+R_S\log\left(1+(e^{(R_0-R_S)/R_S}-1)e^{-t_{\sigma}/R_S}\right)\nonumber\\
     &\approx R_s\left(1+(e^{(R_0-R_S)/R_S}-1)e^{-t_{\sigma}/R_S}\right)
     \label{DSR}
\end{align}
where the position $R_0$ of the shell at $t=0$ is given by  $e^{(R_0-R_S)/R_S}-1 = \frac{(\lambda{}x_0^+)^2}{e}$ and in the second line we have used the late-time (near horizon) limit $t_\sigma\gg{}R_S$. 
Again, we can see that the shell reaches the horizon only as $t_{\sigma}\ra\infty$.

Finally, at early times $t_\sigma\to{}-\infty$ the spacetime is flat up to exponentially small corrections. Indeed, from the Eq. (\ref{DSR}) we find that 
\be
R(t)\approx{}R_0 - t_\sigma 
\label{DSRearly}
\ee
assuming that $R_0\gg{}R_S$ (this apparently holds for sufficiently large shells). The shell moves with the speed of light, the spacetime inside the shell (at $r<R(t)$) is Minkowski while the metric of spacetime outside it (at $r>R(t)$) is given by the Eq. (\ref{DSM}). Since $R(t)\gg{}R_0\gg{}R_S$ at early times, we come to the conclusion presented above.

The overall physical picture of collapse described above can be fully represented by the Penrose diagram on the Fig. \ref{PenroseCGHS} depicting the causal structure of the collapsing classical $(1+1)$-dimensional spacetime.

\subsection{$(3+1)$-dimensional Schwarzschild collapse: classical solution}
\label{4Dclassical}

The case of $(3+1)$-dimensional spherically symmetric classical gravitational collapse of an infinitely thin shell is in many respects similar to the case of the two-dimensional CGHS collapse. This should not be of any surprise: if the metric ${}^4g_{ik}$ of the 4-dimensional collapsing spacetime is represented in the form
\be
{}^{(4)}g_{ik} = g_{ik} + \frac{e^{-2\phi}}{\kappa^2}s_{ik},
\label{Eq:2d4deq}
\ee
where $g_{ik}$ is the metric on the $r-t$ plane, $s_{ik}$ is a unit 2-sphere metric, $\kappa$ is a constant with dimension of inverse length, and angular degrees of freedom are integrated out, the Einstein-Hilbert action of the 4-dimensional problem reduces to the one of two-dimensional dilaton gravity (see, for example, \cite{Ashtekar:2010qz}).\footnote{Strictly speaking, simple equivalence between 2- and 4-dimensional spherically symmetric collapse problems is of course lost at the quantum level. In particular, the form of the one-loop effective action responsible for the trace anomaly is different for the two cases \cite{Balbinot:1998yh,Balbinot2}. Nevertheless, the results which we present below are equally valid for both 2-dimensional and 4-dimensional spherically symmetric collapse problems as should be clear from the procedure of their derivation.}

\begin{figure}[htp]
\includegraphics[width=0.5\textwidth,height=0.3\textheight]{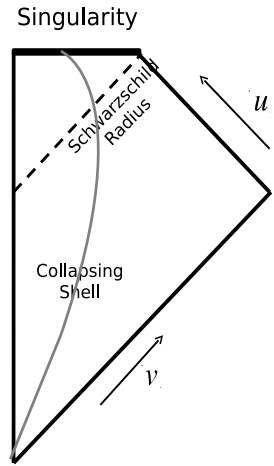}
\caption{Penrose diagram describing 4-dimensional spherically symmetric shell collapsing into Schwarzschild black hole.}
\label{PenroseSchwarzschild}
\end{figure}

The spacetime inside the collapsing spherically symmetric shell of radius $R=R(t)$ is Minkowski due to Birkhoff's theorem:
\be
ds^2 = -dT^2 + dr^2 + r^2 d\Omega^2,
\ee
while outside the shell ($r>R(t)$) it is described by the Schwarzschild metric:
\be
ds^2 = -\!\!\left(1-\frac{R_S}{r}\right)dt^2 +\left(1-\frac{R_S}{r}\right)^{-1}dr^2 +r^2 d\Omega^2.
\ee
The exterior metric has a singularity at $r=0$ covered by the spacelike horizon at $r=R_S$, but the latter is not formed as long as the position of the shell $R(t)>R_S$.
It is well known that from the point of view of a fiducial observer at spatial infinity the $(3+1)$-dimensional gravitational collapse requires an infinite amount of time \cite{Townsend,Misner:1974qy}.

Indeed, let us derive indeed the equations of motion for the shell. To properly match the timelike coordinates inside and outside the shell, one has to refer to the proper time of an observer living on the shell. One finds \cite{Ipser:1983db},\cite{Vachaspati:2006ki}:
\be
\frac{dT}{dt}=\sqrt{B-\frac{1-B}{B}\left(\frac{dR}{dt}\right)^2},
\ee
where $B(t) = 1-\frac{R_S}{R(t)}$.

The equations of motion for the shell are derived from the effective action \cite{Vachaspati:2006ki}
\be
S_{\rm eff} = -4\pi\sigma\int{}dT{}R^2\left(\sqrt{1-\left(\frac{dR}{dT}\right)^2}-2\pi{}G\sigma{}R\right),
\ee
where $\sigma$ is the mass density of the shell. One finds
\be
 \frac{B^{3/2}R^2}{\sqrt{B^2-\left(\frac{dR}{dt}\right)^2}}=\frac{H}{4\pi\sigma(1-2\pi{}G\sigma{}R_S)},
\ee
where $H$ is the conserved (total) mass of the shell. In the near horizon approximation the solution of this equation reads
\be
 R(t)\approx{}R_S+(R_0-R_S)\exp{}(-t/R_S{}).
\label{eq:SchwarschildR}
\ee
The expression (\ref{eq:SchwarschildR}) is close to (\ref{DSR}), the one describing the case of a $(1+1)$-dimensional shell collapsing into CGHS black hole.

It is clear what happens physically at early times $t\to{}-\infty$. The shell moves with the speed of light, and as long as $R(t)\gg{}R_0{}\gg{}R_S$, the spacetime is Minkowski 
up to polynomially small corrections.

Once again, the classical causal structure of the colapsing spacetime can be represented by the Penrose diagram on the Fig. \ref{PenroseSchwarzschild} featuring the event horizon.

\section{CGHS collapse: modes, the action and its diagonalization}
\label{sec:CGHSmodes}

After the brief review of $(1+1)$-D and $(3+1)$-D classical collapse, we are now in the position to discuss quantum effects and their influence on the collapse scenario.

In this Section we consider quantization of a massless scalar field $f$ on the fixed collapsing CGHS background (\ref{DSM}). Equation of motion for the field $f$ reads as usual
\be
\nabla^2f=\frac{1}{\sqrt{-g}}\frac{\partial}{\partial x^{i}}\left(\sqrt{-g}g^{ik}\frac{\partial{}f}{\partial x^{k}}\right)=0,
\label{eq:nabla2}
\ee
where $g_{ik}$ is the metric (\ref{DSM}). The general solution of (\ref{eq:nabla2}) can be represented as a superposition of eigenmodes of the $\nabla^2$ operator:
\be
f(t,r)=\int dk\cdot{}a_{k}(t)f_{k}(r) + {\rm c.c.}
\label{Heisenberg_quant}
\ee
To quantize the field $f$, we then promote the time-dependent amplitudes $a_k(t)$ into quantum mechanical operators. Note that such a prescription is a bit unusual. Instead of the usual procedure of the Schr\"odinger quantization of the field $f(t,r)$, i.e., representing the field as a superposition of \emph{time-dependent} modes
\be
f(t,r)=\int dk\cdot{}a_k{}f_k(t,r)+{\rm c.c.}
\label{Schrodinger_quant}
\ee
with $a_k$, $a_k^\dagger$ being \emph{time-independent} canonical creation and annihilation operators, we cast the time dependence on the amplitudes $a(t)$ themselves.  The main reason for our choice is convenience --- since the collapse problem is intrinsically time-dependent (the position $R(t)$ of the shell is time-dependent and only reaches the Schwarzschild radius at $t\to\infty$ in the slicing which we use), it makes sense to use a formalism where the consequences of this time dependence can be found more easily. The major drawback of our quantization formalism is that the relativistic invariance of observables is less transparent. 

To determine the exact form of the eigenmodes $f_k(r)$ we observe that (a) the modes are naturally divided into left-moving and right moving ones and (b) the form of the modes $f_k(r)$ is different in the regions $r>R(t)$ and $r<R(t)$, where $R(t)$ is the position of the shell. Naturally, the questions of normalization and matching the modes between interior and exterior regions become important. First, it is impossible to match the time-independent modes inside and outside the shell for an arbitrary moment of time $t$ and $r=R(t)$, since one would have to match the coefficients $a_k (t)$ at arbitrary $t$ as well. However, after the quantization is performed, the coefficients $a_k(t)$ carry the information about the occupation numbers for a given mode $f_k(r)$,\footnote{For example, in the WKB regime, when occupation numbers are large, one can roughly estimate $a_k(t)\sim{}\sqrt{n_k}$ times the time-dependent quantum phase.} and these occupation numbers do change with time (particles are produced by the changed background spacetime). Instead, we shall match the modes only at the initial moment of time $t_0$ assuming the vacuum-like initial state for them at $t_0\to{}-\infty$. It is also natural to choose the spacelike hypersurface $R>R_0$ for the normalization of modes (note that if $R_0\gg{}R_S$, the spacetime is nearly flat --- it is flat inside the shell, while all curvature invariants are very small outside the shell, so one would just have a usual Minkowski normalization for the modes).

Taken these considerations into account, we find for a given mode $f_k$ with momentum $k$ (the index $L$ corresponds to left-moving modes, while the index $R$ --- to right-moving ones)
\be
f_{k,L/R}(r)=\frac{1}{\sqrt{2k}}\left(e^{2\lambda r}-e\right)^{\pm\frac{ik}{2\lambda}}
\label{eq:CGHSmodesOUT}
\ee
in the exterior of the shell, i.e., at $r>R(t)$, where $R(t)$ is the position of the shell (\ref{DSR}) and
\be
f_{k,L/R}(r) = \frac{1}{\sqrt{2k}}e^{\pm{}ikr}
\label{eq:CGHSmodesIN}
\ee
inside the shell.

To quantize the matter field $f$, we shall first need the relationship between the internal time, denoted here as $T$ and the external time, denoted $t$. These can be found from the proper time of an observer ``riding'' the shell, denoted $\tau$:
\begin{align}
  \frac{dT}{d\tau}=\sqrt{1+\left(\frac{dR}{d\tau}\right)^2}\nonumber\\
  \frac{dt}{d\tau}=\frac{1}{F}\sqrt{F+\left(\frac{dR}{d\tau}\right)^2}
\end{align}
so that
\be
  \dot{T}=\sqrt{F-\frac{(1-F)}{F}\dot{R}^2}
  \label{TdotSigma}
\ee
where from (\ref{DSM})
\be
  F\equiv 1-e^{1-R/R_s}.
\ee
To find the behavior of $\dot{T}$ as $R_\sigma\to R_S$ we need to know $\dot{R}$. It is easy to show that the solution for $R$ in (\ref{DSR}) results exactly to
\be
  \dot{R}=-F.
  \label{RdotSigma}
\ee
Correspondingly, substituting (\ref{RdotSigma}) into (\ref{TdotSigma}) gives 
$\dot T = F\to0$ as $R\to R_S$.

Now, following \cite{Vachaspati:2006ki}, the action of the scalar field $f$ can be written as the sum of two parts corresponding to the regions inside and outside the shell. One has
\be
S = S_{\rm in}+S_{\rm out},
\label{ActionSum}
\ee
where
\be
S_{\rm in} = \int{}dt\int_0^{R(t)}dr\left(-\frac{(\partial_t{}f)^2}{F}+F(\partial_r{}f)^2\right)
\ee
and
\begin{align}
S_{\rm out} = \int{}dt\int_{R(t)}^\infty{}dr&\left(-\frac{(\partial_t{}f)^2}{1-e^{1-r/R_S}}+\right.\nonumber\\
&\left.+(1-e^{1-r/R_S})(\partial_rf)^2\right).
\end{align}
Let us find out what kind of approximations simplifying the action (\ref{ActionSum}) are possible. 

Since only the time dependent amplitudes $a_k$ are considered quantum mechanical variables, after substituting the expansion (\ref{Heisenberg_quant}) into the action we can explicitly take integrals over $r$. We then find 
\be
S=-\int dt dk dk' \Big{[}\frac{1}{2}K_{kk'}-\frac{1}{2}G_{kk'}\Big{]},
\label{DRadActionMode}
\ee
where the kinetic energy contribution is
\begin{align}
K_{kk'} =&\dot{a}_{k,L}M_{kk'}^{LL,aa}\dot{a}_{k',L} +
\dot{a}_{k,R}M_{kk'}^{RR,aa}\dot{a}_{k',R} + \nonumber\\
&+\dot{a}_{k,R}M_{kk'}^{RL,aa}\dot{a}_{k',L} +
\dot{a}^\dagger_{k,L}M_{kk'}^{LL,ca}\dot{a}_{k',L}+\nonumber\\
&+\dot{a}^\dagger_{k,R}M_{kk'}^{RR,ca}\dot{a}_{k',R} +
\dot{a}^\dagger_{k,R}M_{kk'}^{RL,ca}\dot{a}_{k',L}+\nonumber\\
&+{\rm c.c.},
\end{align}
while the gradient energy contribution is
\begin{align}
G_{kk'} = &{a}_{k,L}N_{kk'}^{LL,aa}a_{k',L} + {a}_{k,R}N_{kk'}^{RR,aa}{a}_{k',R} +\nonumber\\
&+{a}_{k,R}N_{kk'}^{RL,aa}{a}_{k',L}+{a}^\dagger_{k,L}N_{kk'}^{LL,ca}{a}_{k',L}+\nonumber\\
&+{a}^\dagger_{k,R}N_{kk'}^{RR,ca}{a}_{k',R}+{a}^\dagger_{k,R}N_{kk'}^{RL,ca}{a}_{k',L}+\nonumber\\
&+{\rm c.c.}.
\end{align}
The kernels $M_{kk'}$ and $N_{kk'}$ are defined according to
\begin{align}
M_{kk'}^{LL,aa} = & \frac{1}{F(t)}\int_0^{R(t)}{}dr{}f_{k,L}(r)f_{k',L}(r)+\nonumber\\
+&\int_{R(t)}^{+\infty}\frac{dr}{1-e^{1-r/R_S}}{}f_{k,L}(r)f_{k',L}(r),
\label{M1}
\end{align}
\begin{align}
M_{kk'}^{RR,aa} = &\frac{1}{F(t)}\int_0^{R(t)}{}dr{}f_{k,R}(r)f_{k',R}(r)+\nonumber\\
+&\int_{R(t)}^{+\infty}\frac{dr}{1-e^{1-r/R_S}}{}f_{k,R}(r)f_{k',R}(r),
\label{M2}
\end{align}
\begin{align}
M_{kk'}^{LR,aa} = &\frac{1}{F(t)}\int_0^{R(t)}{}dr{}f_{k,L}(r)f_{k',R}(r)+\nonumber\\
+&\int_{R(t)}^{+\infty}\frac{dr}{1-e^{1-r/R_S}}{}f_{k,L}(r)f_{k',R}(r),
\label{M3}
\end{align}
etc.,
\begin{align}
N_{kk'}^{LL,aa} = &\int_{R(t)}^{\infty}{}dr\left(1-e^{1-r/R_S}\right){}f_{k,L}'(r)f_{k',L}'(r)+\nonumber\\
+&F(t)\int_0^{R(t)}{}dr{}f_{k,L}'(r)f_{k',L}'(r),
\label{N1}
\end{align}
\begin{align}
N_{kk'}^{RR,aa} = &\int_{R(t)}^{\infty}{}dr\left(1-e^{1-r/R_S}\right){}f_{k,R}'(r)f_{k',R}'(r)+\nonumber\\
+&F(t)\int_0^{R(t)}{}dr{}f_{k,R}'(r)f_{k',R}'(r),
\label{N2}
\end{align}
\begin{align}
N_{kk'}^{LR,aa} = &\int_{R(t)}^{\infty}{}dr\left(1-e^{1-r/R_S}\right){}f_{k,L}'(r)f_{k',R}'(r)+\nonumber\\
+&F(t)\int_0^{R(t)}{}dr{}f_{k,L}'(r)f_{k',R}'(r),
\label{N3}
\end{align}
etc.

Let us explicitly calculate $(LL,aa^\dagger)$ and $(RR,aa^\dagger)$ matrices. The matrix $M_{kk'}^{LL,aa^\dagger}$ has two contributions --- from the regions inside and outside the shell. The first, interior one, is
\begin{align}
\frac{1}{F(t)}\int_0^{R(t)}{}dr{}f_{k,L}(r)&f_{k',L}^*(r) \nonumber\\
=& \frac{1}{2F(t)\sqrt{kk'}}\int_0^{R(t)}dr{}e^{i(k-k')}\\
=&\frac{H(k-k',R(t))}{2\sqrt{kk'}F(t)},\nonumber
\end{align}
where we have introduced the function
\be
H(k-k',x)=\frac{1}{i(k-k')}\left(e^{i(k-k')x}-1\right).
\ee
The second, exterior, contribution to $M_{kk'}^{LL,aa^\dagger}$ can be in turn expressed as
\begin{align}
\int_{R(t)}^{+\infty}\frac{dr}{1-e^{1-r/R_S}}{}f_{k,L}(r)f_{k',L}^*(r)=
\frac{1}{2k}\delta (k-k')-& \nonumber\\
-\frac{1}{2\sqrt{kk'}}H(k-k',R_S\log{}(e^{R(t)/R_S}-e))&\\
=\frac{1}{2k}\delta (k-k') - \frac{H(k-k',(\lambda{}x_0^+)^2e^{-t/R_S})}{2\sqrt{kk'}}&.\nonumber
\end{align}
Using the properties of the function $H(k-k',x)$, one can further simplify these expressions. As $k-k'\to{}0$ (or $x\to{}0$), the function $H(k-k',x)$ approaches the finite value $1$. At sufficiently large $k-k'$ and $x$ it rapidly oscillates effectively cutting off the correspondent contributions to the action at $|k-k'|\sim{1/x}$. Therefore, integrals of the form $\int dkdk'{}H(k-k',x)A_kA_{k'}^*$ can be estimated as
\begin{align}
\int dkdk'&H(k-k',x)A_kA_{k'}^*\nonumber\\
&=\int{}dkd\Delta{}kH(\Delta{}k,x)A_k{}A_{k-\Delta{}k}^*&\label{Eq:DiagApprox}\\
&\approx{}{}\frac{1}{x}\int{}dk|A_k|^2,\nonumber
\end{align}
where $\Delta{}k=k-k'$ and we used the quasiclassic approximation assuming that the amplitude $A_k$ is a slowly changing function of the momentum $k$.

The Eq. (\ref{Eq:DiagApprox}) implies that the function $H(k-k',x)$ can be safely substituted by the delta function $\delta{}((k-k')x)=\frac{1}{x}\delta{}(k-k')$, when $x$ is sufficiently large. How large is ``sufficiently large''? As we shall see soon, although the ``in'' quantum state of the field $f(t,r)$ (at $t\to-\infty$) is described using momentum quantum numbers $k$, the relevant quantum numbers for the ``out'' state (at $t\to+\infty$) are not $k$, but \emph{exponentially redshifted momenta} $p\sim{}k\cdot\exp{}(t/2R_S)$. Therefore, while the function $H(k-k',x)$  does not completely coincide with $\frac{1}{x}\delta{}(k-k')$ being a smooth non-singular distribution smeared over the interval $|k-k'|\sim{}1/x$ of ``in'' momenta, its width in terms of ``out'' momenta $|p-p'|\sim{}\exp{}(t/2R_S)/x$ vanishes as $t\to{}+\infty$. 

Keeping this in mind, we find that the interior contribution to the matrix $M_{kk'}^{LL,aa^\dagger}$ is approximately given by
\be
\frac{1}{F(t)}\int_0^{R(t)}{}dr{}f_{k,L}(r)f_{k',L}^*(r)\approx\frac{1}{2kF(t)}\delta{}(k-k')
\ee
at early times $t\to-\infty$, when the size of the shell $R(t)$ is very large, and
\be
\frac{1}{F(t)}\int_0^{R(t)}{}dr{}f_{k,L}(r)f_{k',L}^*(r)\approx\frac{R_S}{2\sqrt{kk'}F(t)}
\ee
for $|k-k'|\ll{}1/R(t)\sim{}1/R_S$ at late times $t\to+\infty$. The interior contribution is approximately diagonal for all $k$ and $k'$ at early times. On the other hand, it is only diagonal for $k,k'\gg{}1/R_S$, while all the modes with $k,k'<1/R_S$ mix between each other.\footnote{In fact, the modes with \emph{arbitrary} $k,k'$ mix between each other within the interval $|k-k'|\sim{}1/R_S$ but for $k,k'\gg{}1/R_S$ this interval is small compared to the values of $k$ and $k'$.} We also note that the prefactor $1/F(t)\sim{}1$ at early times $t\to-\infty$ and blows up exponentially at late times $t\to+\infty$.

Similarly, the exterior contribution to the matrix $M_{kk'}^{LL,aa^\dagger}$ is found to be
\be
\int_{R(t)}^{+\infty}\frac{dr}{1-e^{1-r/R_S}}{}f_{k,L}(r)f_{k',L}^*(r)\approx{}\frac{1}{2k}\delta (k-k')
\ee
in the asymptotic future $t\to{}+\infty$, while in the asymptotic past this contribution vanishes entirely. Taking into account the behavior of $F(t)$, we therefore conclude that
\be
M_{kk'}^{LL,aa^\dagger}\approx{}\frac{1}{2k}\delta{}(k-k')
\label{Eq:Mearlytimes}
\ee
at $t\to-\infty$, while at $t\to+\infty$ it is 
\be
M_{kk'}^{LL,aa^\dagger}\approx{}\frac{1}{2kF(t)}\delta{}(k-k')
\label{Eq:Mlatetimes}
\ee
for the modes with momenta $k,k'\gg{}1/R_S$. In principle, these are the modes of the main interest for us as the geometric optics limit can only be applicable for such modes. 
 
A similar analysis of the terms contributing to the matrix $N_{kk'}^{LL,aa^\dagger}$ shows that 
\be
N_{kk'}^{LL,aa^\dagger}\approx{}\frac{k}{2}\delta{}(k-k')
\label{Eq:Nearlylatetimes}
\ee
for both asymptotic past and future infinities $t\to\pm\infty$. 

Corrections to the expressions (\ref{Eq:Mearlytimes}), (\ref{Eq:Mlatetimes}) and (\ref{Eq:Nearlylatetimes}) are suppressed by the parameter $\exp{}(-|t|/R_S)$, very small at $t\to\pm\infty$. At intermediate times $t\to{}0$ these corrections become of the order one, but repeating our analysis and using the properties of the function $H(k-k',x)$ suggests that the only modes affected by the change of background spacetime between $t\to-\infty$ and $t\sim{}0$ are deep IR modes with $k,k'<1/R_0$, where $R_0=R(t=0)\gg{}R_S$.\footnote{As will become clear from the further discussion, this approximation is rather rough but sufficient to address our goals, namely, to determine the structure of the out-vacuum and calculate Bogolyubov coefficients between ``in'' and ``out'' vacua, see Section \ref{Conclusion}.} If we are not interested in those much, we can 
just set the initial quantum state $\Psi(a_k,t=0)$ of the system at $t=0$ as the direct product of the ``in'' (Minkowski) vacuum states for the modes with different $k$.    

Finally, it is worth noting again that at $t\to+\infty$ the matrix $M_{kk'}^{LL,aa^\dagger}$ is not diagonal for the modes with momenta $k,k'<1/R_S$. Therefore, these IR modes with fixed momenta do not diagonalize the action (\ref{ActionSum}) and the associated Hamiltonian. As $t\to+\infty$, these modes redshift away (as we shall see soon, the physical frequency at $t\to+\infty$ is $k\sqrt{F(t)}$) and do not affect the observables.

Applying similar considerations to other terms contributing to the action of the scalar field $f$ and focusing only on the modes with $k\gg{1/R_S}$, we conclude that the action for a given mode can be written approximately as
\be
S = \int dt \left( -\frac{1}{F(t)}\int{}dk\frac{|\dot{a}_k|^2}{k} + \int{}dk\cdot{}k{}|a_k|^2 \right),
\label{eq:S-UV}
\ee
where $F(t)$ is a smooth function with the asymptotic behavior $F(t)\to{}1$ at $t\to-\infty$ and $F(t)\to{}$ at $t\to{}+\infty$. In particular, choosing the function
\be
F(t)= \left\{
\begin{array}{rl}
 1 &\mbox{ when $t<0$} \\
 \frac{(\lambda{}x_0^+)^2}{e}e^{-t/R_S} &\mbox{ when $t\geq{}0$}  
\end{array} \right.
\label{Eq:modelingF}
\ee
is sufficient to capture all the essential physics of the problem --- as should be clear from the previous discussion, changing $F(t)$ while keeping its asymptotic behaviour intact only affects the physics of modes with $k<1/R_0\ll{}1/R_S$. 

The field $a_k$ entering the action (\ref{eq:S-UV}) is not canonically normalized. In order to canonically normalize it, one has to introduce the change of variables
\be
A_k = \frac{a_k}{\sqrt{k}}.
\ee
Correspondingly, the modes (\ref{eq:CGHSmodesOUT}), (\ref{eq:CGHSmodesOUT}) have to be changed to
\be
\tilde{f}_k(r) = \sqrt{k}f_k(r)
\ee
and acquire the form
\be
\tilde{f}_k(r) = \left\{
\begin{array}{rl}
  \frac{1}{\sqrt{2}}e^{\pm{}ikr} &\mbox{ if $r<R_S$} \\
  \frac{1}{\sqrt{2}}\left(e^{2\lambda r}-e\right)^{\pm\frac{ik}{2\lambda}} &\mbox{ if $r>R_S$}
\end{array} \right.
\label{eq:truemodes}
\ee
In what follows we always consider the modes $\tilde{f}_k$ and operators $A_k$, not the modes $f_k$ and operators $a_k$ and shall therefore simply omit the tilde.

Let us derive equations of motion for the quantum mechanical amplitude $A_k(t)$ which follow from the action (\ref{eq:S-UV}). One finds 
\be
\ddot{A_k}-\frac{\dot{F}}{F}\dot{A_k}+k^2{}FA_k=0.
\ee
Introducing the ``scale factor''
\be
b(t)=\frac{1}{\sqrt{F(t)}}=\left\{
\begin{array}{rl}
  1 &\mbox{ if $t<0$} \\
  \frac{e^{1/2}}{\lambda{}x_0^+}e^{t/2R_S} &\mbox{ if $t>0$}
\end{array} \right. 
\ee
we can rewrite this equation as
\be
\ddot{A_k}+2\frac{\dot{b}}{b}A_k+\frac{k^2}{b^2}A_k=0.
\label{Eq:dSmodes1}
\ee
One can immediately come to an important conclusion: \emph{time-dependent amplitudes of the modes of the field $f(t,r)$ in the collapsing CGHS background change as if the field was propagating in the fixed $dS_3$ background}\footnote{More precisely, its planar patch. Moreover, since $b(t)\to{}1$ at $t\to-\infty$, the contracting part of $dS_3$ is absent. Instead, the expanding part of $dS_3$ is smoothly connected to Minkowski spacetime at $t\sim{}0$. It is also important to keep in mind that space-dependent parts of the modes $f_k(r)$ for our background spacetime of course differ from the ones of $dS_3$ (where $f_k(r)$ are plane waves).}
\be
ds^2 = dt^2 - e^{2Ht}(dx^2+dy^2)
\label{dS_3}
\ee
with the horizon size $1/H=2R_S$. Although it might seem surprising, it is not completely unexpected since we know that the ``in'' modes of the field $f$ are subject to \emph{the exponential redshift} when they pass through the collapsing shell.

As usual, instead of solving the Eq. (\ref{Eq:dSmodes1}) it is convenient to deal with the equation of motion rewritten in terms of the ``conformal time''
\be
\eta\approx-\frac{2\lambda{}x_0^+R_S}{e^{t/2R_S+1/2}},
\label{Eq:conformaltime}
\ee
as $t\to+\infty$. For $u_k(\eta)=A_k(\eta)/\sqrt{b(\eta)}$ one finds in the asymptotic future limit 
\be
u_k''+\left(k^2-\frac{3}{4\eta^2}\right)u_k=0.
\label{Eq:umodes}
\ee
This equation can be solved exactly in terms of Bessel functions. The expression for the field $f(t,r)$ is given at arbitrary times $t$ by
\begin{align}
f{}(t,r)=\eta\left( {c}_k{}H_1^{(1)}(k\eta) + {d}_k{}H_1^{(2)}(k\eta) \right)\times{}&\nonumber\\
\times\left(e^{r/R_S}-e\right)^{ikR_S}+{\rm c.c.},&
\label{Eq:phiExact}
\end{align}
where $\eta$ is given by the Eq. (\ref{Eq:conformaltime}), and $c_k$, $d_k$ are arbitrary constants determined by the initial state of the field. 

It is useful to note that the \emph{physical frequency} of the modes, i.e., the one measured by a future asymptotic observer is simply given by
\be
\omega{}=k\eta=k_{\rm phys}.
\label{Eq:physfreq}
\ee

\section{Standard calculation of Bogolyubov coefficients}
\label{sec:CGHSBogolyubov}

The manipulations with the action of the scalar field $f$ performed in the previous Section are rather formal. In order to check whether they have physical sense, we shall now calculate explicitly the Bogolyubov coefficients between in- and out-vacuum states and show that the number of produced particles is given by the Planck distribution if we follow the usual considerations \cite{Hawking:1974sw,Giddings:1992ff}.

At $t\to-\infty$ the spacetime is Minkowski up to exponentially small contributions to the metric. Therefore, we can write for the ``in'' modes
\be
u_k(t,r) = \frac{1}{\sqrt{2k}}e^{-ikt+ikr}.
\label{Eq:CGHSinmodes}
\ee
As for the out-vacuum, in the previous Section we have convinced ourselves that the time-dependent amplitudes $A_k(t)$ change with time as if the scalar field is propagating in the $dS_3$ background spacetime. Correspondigly, it is natural to assume that the out-vacuum of the problem coincides with the Bunch-Davies (Euclidean) vacuum of the scalar quantum field theory in $dS_3$. 
Note that this concerns only the time-dependent part of the field, while the $r$-dependent part is determined by the modes (\ref{eq:CGHSmodesOUT}). Thus, the ``out'' modes are
\be
v_{k'}(t,r) = \frac{1}{2}\sqrt{\pi\eta}H_1^{(2)}(k\eta)(e^{r/R_S}-e)^{ik'R_S},
\label{Eq:CGHSoutmodes}
\ee
where $\eta$ is given by the Eq. (\ref{Eq:conformaltime}).

Since we would like to reproduce the standard answer for the Bogolyubov coefficients and the spectrum of Hawking radiation, we will only be interested at this moment in modes with $k\eta\gg{}1$, the case when geometric optics arguments can be applied. One approximately has for such modes
\begin{align}
v_{k'}(t,r)\approx{}&\frac{1}{\sqrt{2k'}}\left(e^{r/R_S}-e\right)^{ik'R_S}\times\nonumber\\
&\times\exp\left(-2ik'R_S\frac{\lambda{}x^+_0}{e^{1/2}}e^{-t/2R_S}+\frac{3\pi{}i}{4}\right).
\label{Eq:CGHSoutmodesAPP}
\end{align}

To calculate the coefficients of the Bogolyubov transformation between in- and out-vacua, we pick a spacelike hypersurface $t={\rm Const.}\gg{}R_S$ and calculate the scalar products between ``in'' and ``out'' modes on this hypersurface. One finds
\begin{align}
\alpha_{k'k}&=\frac{1}{2\pi}(v_{k'},u_k)&\nonumber\\ 
&\approx{}-\frac{i}{4\pi}\sqrt{\frac{k}{k'}}\int_{R_S}^{+\infty}dr{}e^{-ikx}\left(e^{r/R_S}-e\right)^{ik'R_S}\times&\nonumber\\
&\times\exp\left(-2ik'R_S\frac{\lambda{}x_0^+}{e^{1/2}}e^{-t/2R_S}+\frac{3\pi{}i}{4}+ikt\right).\label{Eq:CGHSBogAlphaCalc}
\end{align}
The integral in the last line can be taken exactly giving
\begin{align}
\alpha_{k'k}=&-\frac{iR_S}{4\pi}\sqrt{\frac{k}{k'}}B(ik'R_S+1,i(k-k')R_S)\times\nonumber\\
\times&\exp\left(-2ik'R_S\frac{\lambda{}x_0^+}{e^{1/2}}e^{-t/2R_S}+\frac{3\pi{}i}{4}+ikt\right),
\label{Eq:CGHSBogAlpha}
\end{align}
where $B(x,y)=\frac{\Gamma{}(x)\Gamma{}(y)}{\Gamma{}(x+y)}$ is the beta-function. 

Similarly, one has
\begin{align}
\beta_{k'k}&=\frac{1}{2\pi}(v_{k'},u_k^*)\nonumber\\ 
&\approx\frac{iR_S}{4\pi}\sqrt{\frac{k}{k'}}B(ik'R_S+1,-i(k+k')R_S)\times\nonumber\\
&\times\exp\left(-2ik'R_S\frac{\lambda{}x_0^+}{e^{-1/2}}e^{-t/2R_S}-i\frac{3\pi}{4}-ikt\right).
\label{Eq:CGHSBogBeta}
\end{align}
It is clear that the relevant\footnote{That is, contributing to the absolute values $|\alpha_{k'k}|$ and $|\beta_{k'k}|$ of Bogolyubov coefficients.} parts of the expressions (\ref{Eq:CGHSBogAlpha}) and (\ref{Eq:CGHSBogBeta}) coincide with the ones presented in \cite{Giddings:1992ff}. Therefore, the analysis of Giddings and Nelson can be completely applied to our case revealing
\be
|\beta_{k'k}|^2=\frac{1}{e^{4\pi{}k'R_S}-1},
\label{Eq:CGHSPlanck}
\ee
the Planck spectrum of Hawking radiation from CGHS collapsing shell. Also, introducing the basis of wavepacket modes \cite{Giddings:1992ff} and tracing out the states localized in the interior of the shell, one finds that the density matrix is completely thermal. 

At this point, we remind ourselves that the two key points of this analysis are (a) using the near horizon approximation (this also provides a bridge to the canonical result \cite{Hawking:1974sw} for (3+1)-dimensional Schwarzschild black holes) and (b) manipulating the contour of integration in  $\alpha_{k'k}$ and $\beta_{k'k}$ in the complex plane. Namely, noting that the main contribution to $\alpha_{k'k}$ and $\beta_{k'k}$ comes from the region near horizon, one finds 
\be
\alpha_{k'k}\sim\int_0^{+\infty}d\Delta{}r{}e^{-ik\Delta{}r}e^{ik'R_S\log{}(\Delta{}r/R_S)},
\label{Eq:CGHSBogNHAlpha}
\ee
and
\be
\beta_{k'k}\sim\int_0^{+\infty}d\Delta{}r{}e^{ik\Delta{}r}e^{ik'R_S\log{}(\Delta{}r/R_S)},
\label{Eq:CGHSBogNHBeta}
\ee
where $\Delta{}r=r-R_S$. The arguments of the exponent in the integrands of (\ref{Eq:CGHSBogNHAlpha}) and (\ref{Eq:CGHSBogNHBeta}) seem to be analytic everywhere in the complex plane of $\Delta{}r$ except the negative real axis. Then, we deform the contour of integration in (\ref{Eq:CGHSBogNHBeta}) to the negative $\Delta{}r$ axis and change the variable of integration as $\Delta{}r\to{}-\Delta{}r$. This gives us the relation 
\be
|\alpha_{k'k}|\approx{}e^{2\pi{}k'R_S}|\beta_{k'k}|,
\ee
which together with the normalization condition for the Bogolyubov coefficients immediately leads to the Planck spectrum (\ref{Eq:CGHSPlanck}).
 
\section{CGHS collapse: quantum kinetics of out-vacuum formation}
\label{sec:CGHSQuantumKineticOUT}

In the calculations of the previous Section it has been assumed that the out-vacuum of the scalar field $f$ coincides with the Bunch-Davies vacuum of $dS_3$. We would like to check now whether this is indeed so, since the Heisenberg-like formalism of quantization (\ref{Heisenberg_quant}) allows us to study the very process of formation of the out-vacuum. 

Our program to address this problem is the following. First, in the Sec. \ref{CGHS:hamiltonian} we derive the functional Schr\"{o}dinger equation for the evolution of the quantum state $\Psi{}(A_k,t)$ of the scalar field $f(t,x)$. This equation describes the first-quantized version of the scalar field theory on the background created by the collapsing shell. Due to the time-dependence built into the problem, the corresponding evolution operator is not diagonal in the basis of ``in'' states. We diagonalize it in the Sec. \ref{sec:QK} and then derive the quantum kinetic equation for the occupation number of states instantaneously diagonalizing the evolution operator. We shall further denote elementary excitations described by these states as ``instantaneous excitations''. In the Sec. \ref{sec:CGHScondensate} we exactly solve this quantum kinetic equation and show that the out-vacuum of the field $f$ is formed as Bose condensate of instantaneous excitations. We then study the properties of the out-vacuum.
 
\subsection{The Hamiltonian and functional Schr\"{o}dinger equation}
\label{CGHS:hamiltonian}

Let us start by deriving functional Schr\"{o}dinger equation for the evolution of the quantum state $\Psi$ of the scalar field $f(t,r)$ \cite{Vachaspati:2006ki}.

Redefining the time variable according to 
\be
  \tau=\int_0^{t}dt' F(t')=\int_0^{t}{}\frac{dt'}{b^2(t')},
  \label{Eta_Sigma}
\ee
we find from the action (\ref{eq:S-UV}) that the Hamiltonian of the field $f$ is given by
\be
 \hat{H} = \frac{1}{2}\int{}dk\left(|A_k|'{}^2+\Omega_k^2{}|A_k|^2\right),
 \label{Eq:Hamiltonian}
\ee
where
\be
  \Omega_k^2(\tau)=\frac{k^2}{F(\tau)}=k^2{}b^2(\tau{})
  \label{omega_sigma}
\ee
and prime denotes derivative with respect to the time $\tau$.

The first quantization of the field theory $f$ on the collapsing CGHS background can be performed by introducing the Fock space of states diagonalizing the Hamiltonian (\ref{Eq:Hamiltonian}). Such states evolve with time according to the functional Schr\"odinger equation:
\begin{align}
  \hat{H}\psi&=\left[-\frac{1}{2}\frac{\partial^2}{\partial ({\rm Re} A_k)^2}+\frac{\Omega_k^2(\tau)}{2}({\rm Re} A_k)^2\right]\psi(A_k,\tau)+\nonumber\\
  &+({\rm Re}\to{\rm Im})=i\frac{\partial\psi(A_k,\tau)}{\partial\tau},
  \label{D_Schrod}
\end{align}
where $A_k$ is as usual the amplitude of the eigenmode with momentum $k\gg{}1/R_S$. Since (\ref{Eq:Hamiltonian}) is simply the Hamiltonian of a harmonic oscillator with a time-dependent frequency, the time-dependent quantum mechanical amplitude $A_k(\tau)$ has a physical meaning of coordinate variable for this oscillator, while $A_k'(\tau)$ has a physical meaning of momentum canonically conjugate to the coordinate variable.

Note that because of the time-dependence of the background one could introduce an infinite family of different Hamiltonians related to each other by time-dependent canonical transformations \cite{Fulling:1979ac}. Among those, we hand-pick the Hamiltonian (\ref{Eq:Hamiltonian}) because of its simplicity and the direct relation to the action (\ref{eq:S-UV}). As will become clear from further discussion, different Hamiltonians related by the time-dependent Bogolyubov transformations of course correspond to different numbers of elementary excitations $N_k(t)$ and different histories of $N_k(t)$, but they provide the very same answer for the final state, out-vacuum, of the scalar field $f(t,r)$ at $t\to\infty$. Therefore, since our only goal is to determine the structure of the out-vacuum state, the choice of $\hat{H}$ from the class of Hamiltonians related to each other by time-dependent canonical transformations is largely arbitrary.

\subsection{Deriving the quantum kinetic equation}
\label{sec:QK}

While it is possible to solve this time-dependent Schr\"odinger equation directly \cite{Vachaspati:2006ki,Greenwood:2010sx}, 
we shall develop below a different, more physically transparent approach allowing us to find an explicit expression for the occupation numbers $N_k$. Namely, let us consider Ehrenfest equations of motion for the time-dependent operators $A_k(\tau)$ following from the Hamiltonian (\ref{Eq:Hamiltonian}):
\be
  \frac{dA_k}{d\tau}=p_{k}, \hspace{2mm} \text{and} \hspace{2mm} \frac{dp_{k}}{d\tau}=-\Omega^2(\tau) A_k.
  \label{H_eq}
\ee
Instead of $A_k$ and $p_k$ let us introduce another pair of operators
\begin{align}
  \hat{b}_k&=\frac{1}{\sqrt{2\Omega(\tau)}}\left(p_k-i\Omega(\tau) A_k\right) \hspace{2mm} \text{and}\nonumber\\
  \hat{b}_k^{\dagger}&=\frac{1}{\sqrt{2\Omega(\tau)}}\left(p_k+i\Omega(\tau) A_k\right)
  \label{a_hat}
\end{align}
It is clear that the operators $\hat b$ and $\hat b^\dagger$ are in the same relation to the operators $A_k$ and $p_k = -i\partial/\partial{}A_k$ as the creation and annihilation operators to the operators of momentum and coordinate in the case of a single quantum mechanical harmonic oscillator. These $b$ operators \emph{diagonalize the instantaneous Hamiltonian} (\ref{Eq:Hamiltonian}) for all values of the time dependent angular frequency $\Omega(\tau)$, and the combination $\hat b\hat b^\dagger$ gives the corresponding occupation number for a mode with momentum $k$.  

By explicitly taking derivatives of (\ref{a_hat}), one finds equations of motion for the operators $\hat{b}$:
\begin{align}
  \frac{d\hat{b}}{d\tau}&=-i\Omega\hat{b}-\frac{\Omega'}{2\Omega}\hat{b}^{\dagger},\nonumber\\
  \frac{d\hat{b}^{\dagger}}{d\tau}&=i\Omega\hat{b}^{\dagger}-\frac{\Omega'}{2\Omega}\hat{b}.
  \label{a_hat_der}
\end{align}
Note that we have additional anomalous terms in the R.H.S.'s of the Eqs. (\ref{a_hat_der}) due to the time dependence of the frequency $\Omega$. 

Now, if we define the occupation number as $N_k=\hat{b}_k\hat{b}_k^{\dagger}$ and an anomalous operator
$\chi{}_k^*=\hat{b}_k^\dagger\hat{b}_{-k}^\dagger$ (the latter describes pair creation and is not zero due to the time dependence of $\Omega$), and using (\ref{a_hat_der}) we come to the three coupled first order differential equations
\begin{align}
  \frac{dN_k}{d\tau}&=-\frac{\Omega'}{2\Omega}\left(\chi_k^{\dagger}+\chi_k\right)\label{N_D}\\
  \frac{d\chi_k}{d\tau}&=-2i\Omega\chi_k-\frac{\Omega'}{2\Omega}\left(2N_k+1\right)\label{sigma}\\
  \frac{d\chi_k^{\dagger}}{d\tau}&=+2i\Omega\chi_k^{\dagger}-\frac{\Omega'}{2\Omega}\left(2N_k+1\right).\label{sigma_d}
\end{align}
where primes denote derivatives with respect to $\tau$. Instead of $\chi_k$ and $\chi^*_k$ it is more convenient to use the operators $X_k$ and $Y_k$ defined according to
\begin{align}
  X_k&=\chi_k+\chi_k^{\dagger},\nonumber\\
  Y_k&=i(\chi_k-\chi_k^{\dagger})\label{XYdef}
\end{align}
In terms of them, we can rewrite (\ref{N_D}) - (\ref{sigma_d}) as
\begin{align}
  \frac{d N_k}{d\tau}&=-\frac{\Omega'}{2\Omega}X_k\label{N_X}\\
  \frac{d X_k}{d\tau}&=2\Omega Y_k-\frac{\Omega'}{\Omega}\left(2N_k+1\right)\label{X}\\
  \frac{d Y_k}{d\tau}&=-2\Omega X_k. \label{Y}
\end{align}

Here we note that (\ref{D_Schrod}) is written in terms of $\tau$, not asymptotic observer
time $t$. Hence, the frequency $\Omega$ is not the physical frequency. From (\ref{Eta_Sigma}) we see that in respect to time of observer living at spatial infinity, we can write the physical frequency as
\be
  \omega{}(t)=k\sqrt{F(t)}=\frac{k}{b(t)}.
  \label{omega_t}
\ee
Using (\ref{Eta_Schwarz}) and (\ref{omega_t}) we can also rewrite (\ref{N_X}) - (\ref{Y}) in terms of the observer time $t$ as
\begin{align}
  \dot{N}_k&=\frac{\dot{\omega}}{2\omega}X_k\label{N_t},\\
  \dot{X}_k&=2\omega Y_k+\frac{\dot{\omega}}{\omega}\left(2N_k+1\right)\label{X_t},\\
  \dot{Y}_k&=-2\omega X_k \label{Y_t}.
\end{align}

Let us briefly discuss the physics behind the system of equations (\ref{N_t}) - (\ref{Y_t}). The expectation value\footnote{Recall that we use the Heisenberg-like picture where states are time-independent, while the operators are time-dependent.} of the operator $N_k$ is related to the occupation number for the mode with a given momentum $k$. If the frequency $\omega$ does not depend on time, the system (\ref{N_t}) - (\ref{Y_t}) reduces to
\begin{align}
  \dot{N}_k&=0,\nonumber\\
  \dot{X}_k&=2\omega Y_k,\nonumber\\
  \dot{Y}_k&=-2\omega X_k \label{Y_tstatic},
\end{align}
i.e., the occupation number remains the same while the expectation values of the anomalous operators $X_k$ and $Y_k$ oscillate with time. This is exactly what one should expect, since the time dependence of the Heisenberg operator $\hat b_k(t)$ in the static situation is simply determined by the phase factors $\exp (-i\hat{H}t)$, and the anomalous operators $X_k$ and $Y_k$ are trivially related to the Heisenberg operators $\hat b_k$ and $\hat b_k^\dagger$ (see (\ref{XYdef})). When the frequency $\omega$ becomes time-dependent,  spontaneous particle production from the vacuum is possible, with the amplitude of the corresponding process being proportional to the adiabaticity parameter $\dot{\omega}/{\omega}$. This spontaneous particle production is the reason why the anomalous terms appear on the right hand side of (\ref{N_t}) - (\ref{Y_t}).

The expectation values written in terms of the ``in'' momenta $k$ are of little interest for an observer located at spatial infinity who sees the rays passing through the collapsing shell, as in the original picture considered by Hawking. The frequencies of these rays are exponentially strongly redshifted according to (\ref{omega_t}), so in order to construct the map between our Heisenberg-like picture and the standard calculation of Bogolyubov coefficients, we have to rewrite our observables as functions of time $t$ measured by a clock at spatial infinity and the physical frequency $\omega{}(t)$ given by \ref{omega_t}). 

It is easy to check that the operators $N_k$, $X_k$ and $Y_k$ do not commute with each other \cite{NXYnocomm}. What this fact really tells us is that for a given observer it is never possible to measure simultaneously the occupation number for the given mode and the corresponding quantum phase of the state. Also, it is impossible to prepare a state with a fixed occupation number and a fixed value of the quantum phase: we can either prepare a Fock state with the fixed $N$ but undetermined phase or a correlated state with a fixed value of the quantum phase which is a superposition of Fock states with fixed $N_k$'s. However, what the equations (\ref{N_t}) - (\ref{Y_t}) want to tell us is that the pairs of quanta of ``instantaneous excitations'' are produced in a correlated state (namely, the quantum phases of both particles in the pair are correlated), and this correlation remains preserved with time. In fact, it is possible to exclude $X_k$ and $Y_k$ from the system (\ref{N_t}) - (\ref{Y_t}) writing a single integro-differential equation for the operator $N_k$. It has the form
\begin{align}
\frac{dN_k(t)}{dt}=\frac{\dot{\omega}(t)}{\omega{}(t)}\int_0^t{}dt'{}\frac{\dot{\omega}(t')}{2\omega{}(t')}(2N_k(t')+1)\times\nonumber\\
\times\cos(2[\Theta{}(t)-\Theta{}(t')]),
\label{eq:kinetic_eq}
\end{align}
where
\be
\Theta{}(t)-\Theta{}(t')=\int_{t'}^t{}dt''\omega{}(t'').
\label{eq:phases}
\ee

This integro-differential equation is nothing but a quantum kinetic equation describing production of instantaneous excitations on the collapsing background. Keeping in mind that it is equivalent to the system of three linear differential equations (\ref{N_t}) - (\ref{Y_t}), we are now ready to study the solutions of (\ref{eq:kinetic_eq}). 

\subsection{``Out'' vacuum as a condensate of instantaneous excitations}
\label{sec:CGHScondensate}

Since the Hamiltonian equations of motion for the mode amplitude $A_k(\eta)$ following from the Hamiltonian (\ref{Eq:Hamiltonian}) can be solved exactly, so can the quantum kinetic equation (\ref{eq:kinetic_eq}) for the occupation number $N_k(t)$ of the instantaneous modes. If we introduce the new variable $z = 2\omega{}(t)R_S$, the exact solution reads  
\begin{equation}
N(t,z) = \frac{k^2R_S{}L}{z}\left(\left|\frac{dA_k}{dz}\right|^2+|A_k|^2\right)-\frac{1}{2},
\label{Eq:QKexactSol}
\end{equation}
where $L$ is the linear size (``volume'') of the system\footnote{
One can understand its origin as follows. The Hamiltonian of the field is given by (\ref{Eq:Hamiltonian}). Calculating the energy expectation value, one has
$$
E=\langle\hat{H}\rangle=\delta{}(0)\int{}dk\Omega_k\left(N_k+\frac{1}{2}\right),
$$
where the delta function appears due to its presence in the relation $\langle{}b_kb^\dagger_{k'}\rangle=N_k\delta{}(k-k')$. Its dimension is inverse energy and the physical meaning is the linear size of the system, as becomes manifest when the discrete normalization for time-independent modes $f_k(r)$ is used.
} 
which we shall omit further,
\begin{equation}
A_k = a_k{}z{}H^{(1)}_1(z)+b_k{}z{}H^{(2)}_1(z)
\label{Eq:ASol}
\end{equation}
(see also Eq. (\ref{Eq:phiExact})) and
\begin{align}
&\frac{A_k(z)}{dz} = a_k{}{}H^{(1)}_1(z)+b_k{}{}H^{(2)}_1(z)+\\ \label{Eq:APrimeSol}
&+ \frac{z}{2}\left(a_k(H_0^{(1)}(z)-H_2^{(1)}(z))+b_k(H_0^{(2)}(z)-H_2^{(2)}(z))\right).&\nonumber
\end{align}
The time dependence of the occupation number here is implicit: while the constants $a_k$ and $b_k$ are  $z$-independent, they still depend on the value of ``in'' momentum $k$, because their values are determined by the initial condition for the occupation numbers $N_k$ at $t=0$. Substituting $k = z\exp{}(t/2R_S)/2R_S$, one finds the actual dependence of $N_k$ on time.

The vacuum-like initial conditions $N_k(t=0)=0$ for the quantum kinetic equation (\ref{eq:kinetic_eq}) can be set as follows. In principle, we are only interested in the late-time behavior of $N_k(t,z)$ at physical frequencies $\omega\gg{1/R_S}$, i.e., $z\gg{}1$. Apparently, large physical, ``out'', momenta correspond to even larger ``in'' momenta since
\begin{equation}
1\ll{}z=2\omega{}R_s=2kR_se^{-t/2R_S}\ll{}2kR_S.
\end{equation}
Therefore, to fix the values of $a_k$ and $b_k$ by the initial conditions at $t\to{}0$, one can safely use the expansion of Hankel functions in (\ref{Eq:QKexactSol}) at large values of argument. Neglecting rapidly oscillating factors, one then finds at $t\to{}0$
\begin{equation}
N_k(t=0)\equiv{}N_{0,k}\approx{}\frac{4k^2R_S}{\pi}\left(|a_k|^2+|b_k|^2\right)-\frac{1}{2},
\label{Eq:QKIC1}
\end{equation}
i.e.,
\begin{equation}
|a_k|^2+|b_k|^2 = \frac{\pi}{4k^2{}R_S}\left(N_{0,k}+\frac{1}{2}\right),
\label{Eq:QKIC}
\end{equation}
in particular, for the vacuum-like initial condition one has
\begin{equation}
|a_k|^2+|b_k|^2 = \frac{\pi}{8k^2{}R_S}.
\label{Eq:QKICVac}
\end{equation}
As we see, to the leading order in $1/k$, the complex coefficients $a_k$ and $b_k$ cannot be separately fixed by the initial condition $N_k(t=0)=N_{0,k}$, only the combination $|a_k|^2+|b_k|^2$ can. 

Let us first consider the case of vacuum-like initial conditions $N_{0,k}=0$.
Apparently, the pair $a_k=0$, $b_k = \sqrt{\pi/{8k^2{}R_S}}$ which belongs to the set defined by the Eq. (\ref{Eq:QKICVac}) corresponds to the Bunch-Davies out-vacuum of the field $f(t,r)$. Indeed, using (\ref{Eq:ASol}) one finds
\begin{equation}
A_k=b_kH^{(2)}_1(z)=\frac{\sqrt{\pi\eta}}{2\sqrt{b(\eta)}}H_1^{(2)}(k\eta),
\label{Eq:QKBunchDavies}
\end{equation}
and the ``out'' modes of the field $f(t,r)$ coincide with (\ref{Eq:CGHSoutmodes}). Generally, for arbitrary $a_k$ and $b_k$ satisfying (\ref{Eq:QKICVac}), the time-independent asymptotics of the amplitude $A_k$ is given by
\begin{equation}
A_k = \frac{\sqrt{\pi\eta}\left(H_1^{(2)}(k\eta)+
e^{\alpha_k}H^{(1)}_1(k\eta)\right)}{\sqrt{b(\eta)(1-e^{\alpha_k+\alpha_k^*})}},
\label{Eq:QKAllenMottola}
\end{equation}
where ${\rm Re}\alpha_k<0$ and $b(\eta)=-4R_S^2/\eta^2$ is the ``scale factor''. When the constant parameters $\alpha_k=\alpha$ are $k$-independent, the Eq. (\ref{Eq:QKAllenMottola}) describes Mottola-Allen $E(2)$-invariant vacua of $dS_3$ \cite{Mottola:1984ar,Allen:1985ux}. The general case of $\alpha_k$'s depending on $k$ corresponds to arbitrary vacua of three-dimensional de Sitter space, not necessarily invariant with respect to any subgroups of the symmetry group $SO(1,3)$ of $dS_3$. Apparently, asymptotic states developed from non-vacuum like initial conditions $N_{0,k}\ne{}0$ belong to Fock spaces built on top of generalized Mottola-Allen vacua.  

As follows from (\ref{Eq:QKexactSol}), the spectrum of instantaneous excitations constituting the out-vacuum approaches the time-independent asymptotics at $t\to+\infty$. It behaves as a power law at $z\gg{}1$:
\begin{equation}
N(\omega)\sim{}\frac{1}{(2\omega{}R_S)^2}+O((2\omega{}R_S)^{-3}).
\label{Eq:QKPowerLaw}
\end{equation}
The behavior at low frequencies $z\to{}0$ (in particular, the behavior of the zero mode $k=0$) seems to be singular, but in reality this singularity is cured by the time dependence. We shall now show this explicitly.  

\begin{figure}[htp]
\includegraphics{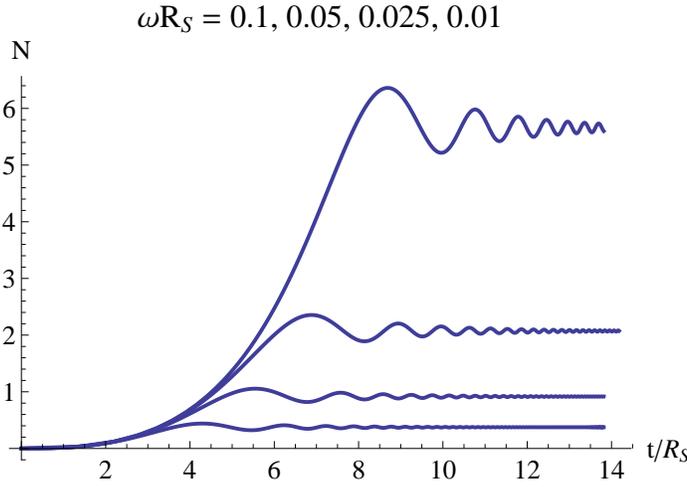}
\caption{Occupation numbers $N$ of instantaneous excitations versus $t/R_S$ for various values of the physical frequency $\omega^{(t)}R_S$ in the CGHS collapsing setup. Here the curves are higher for lower values of $\omega^{(t)}R_s$ in accordance with (\ref{omega_t}). After a finite time of the order $R_S$, the occupation number reaches a plateau and remains constant, up to diminishing fluctuations. This can be understood as the occupation increasing exponentially after some finite time, while the physical frequency $\omega^{(t)}$ is exponentially redshifted, causing the number of particles in a particular mode to become constant over time. This corresponds to the formation of out-vacuum of the scalar field $f(t,r)$.}
\label{DNvst}
\end{figure}

Instead of using exact solution (\ref{Eq:QKexactSol}) it is useful to determine the approximate behavior of $N_k$ at low frequencies directly from the quantum kinetic equation. The latter is a power tool for analyzing particle production and vacuum polarization in QFTs on curved background spacetimes, and this will gain one some intuition regarding behavior of the solutions of (\ref{eq:kinetic_eq}).
Let us start with 
the zero mode $\omega=0$. Taking the derivative of (\ref{X_t}) and noting that
\bd
  \frac{\ddot{\omega}}{\omega}\approx\left(\frac{\dot{\omega}}{\omega}\right)^2\approx\frac{1}{4R_s^2}
  \label{freq_late}
\ed
at late times $t\gg{}1/R_S$, we can write $\ddot{X}$ as
\bd
  \ddot{X}=\frac{X}{4R_s^2},
\ed
where we used (\ref{N_t}). This has the solution
\be
  X\approx \frac{1}{2}e^{-t/2R_s}\left(1-e^{t/R_s}\right),
  \label{X_mid}
\ee
which gives for $N$
\be
  N\approx\frac{1}{4}e^{-t/2R_s}\left(e^{t/2R_s}-1\right)^2,
  \label{N_mid}
\ee
or as $t\ra\infty$
\begin{align}
  N&\approx\frac{1}{4}e^{t/2R_s},\label{N_late}\\
  X&\approx-\frac{1}{2}e^{t/2R_s}\label{X_late}.
\end{align}

We conclude that the occupation number of the zero mode of instantaneous excitations grows exponentially with time.  
In fact, as clear from the plot, the spectrum $N_k$ remains time dependent for a finite interval of frequencies, but this interval shrinks to zero with time.  Let us now show how exactly it happens and what is the $\omega$ dependence of the infrared time dependent part of the spectrum. Note that although the fixed momentum states with $k\lesssim{}R_S^{-1}$ we calculate the occupation numbers for are not the true eigenstates of the Hamiltonian, they do provide the states that a fiducial observer will measure at asymptotic infinity.

\begin{figure}[t]
\includegraphics{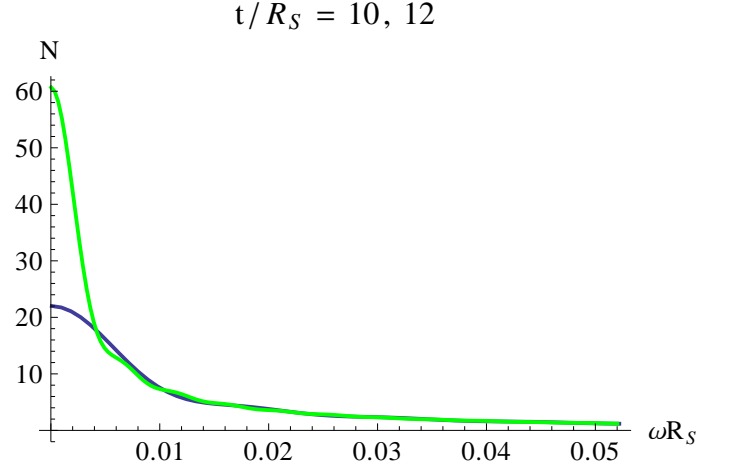}
\caption{The spectrum of instantaneous excitations $N$ versus $\omega^{(t)}R_S$ for $t/R_S=10$ (blue) and $t/R_S=12$ (green) in the CGHS collapsing setup.}
\label{DNvsw}
\end{figure}

The spectrum does appear to become more singular as $t/R_S$ increases. This behavior can again
be seen from (\ref{N_t}) - (\ref{X_t}), and at $\omega\to{}0$ one finds 
\begin{align}
  2N\approx&2R_s\alpha\omega e^{-t/2R_s}+\frac{\gamma}{2R_s\omega}e^{t/2R_s}\nonumber\\
       \approx&\frac{\gamma}{2R_s\omega}e^{t/2R_s}\label{eq:IRspectrum}\\
  X=&2R_s\alpha\omega e^{-t/2R_s}-\frac{\gamma}{2R_s\omega}e^{t/2R_s}\nonumber\\
       \approx&-\frac{\gamma}{2R_s\omega}e^{t/2R_s},
\end{align}
where $\alpha$ and $\gamma$ are constants of integration. The dependence on $\omega^{-1}$ is explicit.  

The expression (\ref{eq:IRspectrum}) is only valid for the interval of $\omega\lesssim{}R_S^{-1}\exp{}(-t/2R_S)$. It is very easy to understand the behaviour of the IR cutoff $\omega_{\rm IR}$ separating IR time-dependent part of the spectrum from its ultra-violet (UV) time-independent part. The $\omega=0$ is the fastest growing mode as all modes with $\omega>0$ have lower occupation numbers. 
Comparing behaviour of the stationary distribution $N(\omega)$ at low frequencies to the time-dependent behaviour of the zero mode, we conclude that $\omega_{\rm IR}\sim{}\frac{1}{{}R_S}\exp{}(-t/2R_S)$.

\section{Quantum kinetics of out-vacuum formation for $3+1$-dimensional Schwarzschild collapse}
\label{sec:MSF}

Let us now consider the case of $(3+1)$-dimensional spherically symmetric collapsing shell. Many of our conclusions applicable to the CGHS collapse hold for this case as well.

To determine the spectrum of instantaneous excitations and describe the quantum kinetics of out-vacuum formation, we first briefly review the set-up and Hamiltonian for the radiation coupled to the background of a collapsing massive shell.

Once again, the action of the scalar field takes on different forms,
one for the interior and one for the exterior of the shell. Near the horizon, where most of the
particle production will occur, the dominant parts of the interior and exterior actions are used to
determine the overall action and the Hamiltonian.

Following \cite{Vachaspati:2006ki} in choosing the real basis of modes $f_k(r)$ for the massless\footnote{The case of massive scalar field is considered in \cite{Greenwood:2010sx}.} scalar field $f$ instead of the complex basis analogous to (\ref{eq:CGHSmodesOUT}), (\ref{eq:CGHSmodesIN}) and neglecting contributions to the action exponentially subdominant at $t\to{}+\infty$, one finds similar to the case of CGHS collapse:
\begin{align}
  S\approx\int dt\sum_{k,k'}\Big{[}&-\frac{1}{B}\dot{A_{k'}}M_{kk'}\dot{A_k}\nonumber\\
       &+A_{k'}N_{kk'}A_k\Big{]},
   \label{4DRadAction}
\end{align}
where
\be
M_{kk'}=\frac{1}{2\pi}\int_0^{R_S}dr{}r^2f_k(r)f_{k'}(r),
\ee
\be
N_{kk'}=\frac{1}{2\pi}\int_{R_S}^\infty\left(1-\frac{R_S}{r}\right)f'_k(r)f'_{k'}(r).
\ee
Again, matrices $M$ and $N$ are almost diagonal when both $k,k'\gg{}R_S^{-1}$, while infrared modes (with $kR_s\lesssim{}1$) are getting strongly mixed since they do not diagonalize the Hamiltonian of the field $f$. The Hamiltonian for a given eigenmode $B_k$ is
\be
  \hat{H}=-\frac{1}{2m}\frac{\partial^2}{\partial B^2}+\frac{1}{2}m\Omega^2(\tau)B^2
  \label{H_eta}
\ee
where $m$ is the corresponding eigenvalue of the matrix $M$,
and we have defined the ``conformal'' angular frequency as
\be
  \Omega^2(\tau)\equiv\frac{\Omega_0^2}{B}=\frac{K}{mB},
  \label{omega_sq_Schwarz}
\ee
and $K$ is the eigenvalue of the matrix $N$. The conformal time is given by
\be
  \tau=\int dt'B(t')
  \label{Eta_Schwarz}
\ee
where
\be
  B\equiv1-\frac{R_S}{R(t)}
  \label{B}
\ee
and
\be
R(t)\approx R_S+(R_0-R_S)e^{-t/R_S}
\ee
is the position of the shell. This is again just the Hamiltonian for uncoupled harmonic
oscillators with time-dependent frequencies.

The spectrum of instantaneous radiation constituting the out-vacuum of the scalar field can be found by solving the quantum kinetic equation. Performing the analysis analogous to the one presented in the previous sections, we find that the quantum kinetic equation describing production of quanta of the instantaneous excitations is reduced to
the set of three coupled first order differential equations (\ref{N_t})-(\ref{Y_t}) with
\be
\omega (t) = \Omega_0\sqrt{B(t)}=\frac{\Omega_0}{a(t)},
\ee
where $a(t)\sim{}\exp{}(t/2R_S)$ is the ``scale factor'' of $dS_3$ space, which the modes $A_k$ effectively propagate in. Correspondingly, solution of these equations shows that all the conclusions of the previous sections remain the same:
\begin{itemize}
\item the spectrum of instantaneous excitations quickly (within the time scale of the order $t\sim{}R_S^{-1}$) approaches the time-independent asymptotics (\ref{Eq:QKexactSol}); the form of this asymptotics determines the out-vacuum state of the scalar field $f(t,r)$,
\item the form of the most general out-vacuum state is given by the Eq. (\ref{Eq:QKAllenMottola}),
\item the UV part $\omega{}R_S\gg{}1$ of the spectrum of instantaneous excitations behaves as a power law (\ref{Eq:QKPowerLaw}),
\item the deep infrared part $\omega{}R_S\ll{}\exp{}(-t/2R_S)$ of the spectrum of instantaneous excitations is not singular, instead the occupation numbers there keep growing exponentially with time;
also, at very small frequencies the fixed momentum states are not the ones diagonalizing the Hamiltonian (\ref{H_eta}).
\end{itemize}

\section{Review of results and discussion}
\label{Conclusion}

In the present paper we study the gravitational collapse of an infinitely thin shell in $(1+1)$-dimensional dilaton gravity (see the classic paper \cite{Callan:1992rs}) as well as a spherically symmetric collapse in $(3+1)$-dimensional gravity. For both setups, the Schwarzschild-like coordinate system is used where full collapse takes infinite time, but the shell approaches its apparent horizon exponentially quickly.

Following Hawking \cite{Hawking:1974sw} (and mostly using the same approximations --- the limit of geometrical optics, near-horizon approximation, analyticity of observables in the complex plane of ``out'' momentum everywhere except the negative real axis, etc.), we consider the quantum field theory of a massless scalar field on a fixed background spacetime corresponding to a collapsing shell. Quantum radiation is emitted during such collapse prior to the formation of the event horizon. This leads to the ``polarization'' of the in-vacuum of the scalar field $f(t,r)$ (i.e. to the creation of a coherent state of excitations of the in-vacuum) and to Hawking evaporation of the collapsing shell. In this paper, we are especially interested in understanding the dynamics of the vacuum polarization, although the computation of the spectrum of Hawking radiation is also performed. Our ultimate goal is to better understand the nature of the information loss paradox. The method which we develop and use to study dynamics of the vacuum polarization, deriving and solving the quantum kinetic equation for particle production during collapse, has to our knowledge never been used in this context in the literature. 

The main result of our paper is the following: it is shown that within a time scale of the order $R_S$ the out-vacuum of the field $f(t,r)$ is formed as a Bose-Einstein condensate (BEC) of ``in'' excitations. Such a condensate is associated with the out-vacuum because it coincides with the lowest energy state as $t\to{}+\infty$. Contrary to the standard assumption about the form of the out-vacuum, the BEC can become inhomogeneous depending on the history of collapse. Furthermore, the information about this collapse history remains encoded in the quantum phase(s) of the BEC, which in principle can be measurable as we discuss below.  
\subsection{Main points of the paper}
\label{sec:mainpoints}
As the way we come to this conclusion is rather technical, it is useful to list and discuss the main steps. \smallskip
\newline 1. Heisenberg-like quantization of the scalar field $f(t,r)$ is used instead of the Schrodinger quantiztion 
(Eq. (\ref{Schrodinger_quant}) usually found in literature: the field is expanded into space-dependent modes $f_k(r)$, while the time-dependent amplitudes $A_k(t)$ of the modes are promoted into quantum mechanical operators (see (\ref{Heisenberg_quant}). This allows one to explicitly follow the time dependence of observables that are the functions of the operators $A_k(t)$, such as the expectation value of the stress energy tensor of the field $f$, and its two-point and multiple-point correlation functions. We do not explicitly separate the Hilbert space $\cal{H}$ of the problem into interior and exterior regions ${\cal H}_i$ and ${\cal H}_e$. Although the analytic form of the space-dependent modes $f_k$ is different inside and outside the shell, the quantum mechanical amplitude of the mode $A_k(t)$ is a single variable. Consequently,  during the collapse evolution  $\cal{H}$ never becomes the direct product of ${\cal H}_i$ and ${\cal H}_e$ in the slicing that we use. Not even as $t\to+\infty$. This is an additional argument in favor of unitarity preservation. Also, note that this Heisenberg quantization picture is very much in the spirit of black hole complementarity conjecture \cite{Balasubramanian:2011dm}: the time dependent amplitudes $A_k{}(t)$ ``know'' about both the interior and the exterior of the collapsing shell.
\newline 2. The action of the scalar field $f(t,r)$ receives contributions from both the interior and the exterior of the collapsing shell (see (\ref{ActionSum})). We show that some of the contributions to the action can be neglected in the limit $t\to+\infty$ compared to others. Roughly speaking, the kinetic contribution to the action of the scalar field as a functional of the amplitudes $A_k(t)$ is dominated by the region inside the shell, while the gradient contribution is dominated by the region outside the shell. Of course, when we neglect some pieces of the action compared to  others, we explicitly violate the locality of the theory. Yet, this is not  of great concern, as it remains clear how to compute local quantities such as the expectation value of the stress-energy tensor using the formalism that we develop. Also, it was long believed that the ultimate resolution of the information loss paradox is related to taking non-local effects into account (see for example  \cite{Giddings:2009ae}). We strongly feel that the developed formalism captures the leading such effect.
\newline 3. Once only leading (in the late-time limit $t\to+\infty$) terms are retained, the action is shown to be equivalent to the action of a massless scalar field on a $dS_3$ background (see (\ref{Eq:dSmodes1}) and following discussion). This is not terribly surprising.\footnote{What perhaps {\em is} surprising is the effective dimensionality of this de Sitter space; note that the original collapse problem is two-dimensional.} In the limit $t\to{}+\infty$, the collapsing shell is located exponentially close to its horizon, and the modes of the scalar field $f(t,r)$ passing through the shell experience exponential redshift. Since the shell is in motion, the amount of redshift itself depends on time. It should therefore be possible to quantify the effect of redshift by introducing the scale factor $b(t)$ relating the frequencies of ``in'' and ``out'' modes. The developed formalism allows us to do exactly that. 
\newline 4. As spacetime is flat at $t\to-\infty$, up to exponentially small corrections to the Minkowski metric in the case of CGHS collapse, we assume that at $t=0$ the quantum state of the scalar field remains in the ``in'' (Minkowski) vacuum  for  modes with momenta $k>R_0$ (where $R_0=R(t=0)$ is the radius of the shell at $t=0$). In a sense, the time-dependent amplitudes $A_k(t)$ of the modes effectively propagate in a spacetime with scale factor $b(t)$, which behaves exponentially at $t\to+\infty$ and approaches $1$  as $t\to{}-\infty$, so the effective spacetime is not $dS_3$, but the expanding planar patch of $dS_3$ smoothly connected to the flat Minkowski space at $t\sim{}0$.
We correspondingly set vacuum-like initial conditions for the modes with $k\gg{}R_0^{-1}$ of quantum state of the scalar field at $t=0$. 
\newline 5. Assuming that the out-vacuum of the field $f$ is effectively the Bunch-Davies (``Euclidean'') vacuum for the scalar QFT on $dS_3$ background, we compute the Bogolyubov coefficients between ``in'' and ``out'' vacua in Sec. \ref{sec:CGHSBogolyubov} for the CGHS collapse and show that they precisely coincide with the standard result by Giddings and Nelson \cite{Giddings:1992ff}, and the Hawking radiation is expected to have the Planck spectrum. It is also clear from the derivation that the same result applies for the case of $(3+1)$-dimensional spherically symmetric collapse of a thin shell.
\newline 6. Instead of assuming the explicit form of the out-vacuum, we would like to derive this form explicitly as a result of the evolution of the in-vacuum polarization. To describe this evolution, we use the formalism of the quantum kinetic equation developed in Sec. \ref{sec:QK}. This formalism was first introduced in \cite{SZ} and mostly used to study particle production in the cosmological context (see for example recent papers \cite{MottolaQK}). Historically, the formalism of the quantum kinetic equation fell out of favor due to the difficulties in prescribing the physical content to the definition of particles for QFTs in curved spacetimes and likewise to the procedure of instantaneous diagonalization of the Hamiltonian \cite{Fulling:1979ac}. In the next subsection we explain how to deal with these difficulties and understand them, and how to extract the physical content from the quantum kinetic equation describing particle production in a QFT on a fixed curved background.
\newline 7. Solving the quantum kinetic equation exactly in Sec. \ref{sec:CGHScondensate}, we find that, within a time scale $t\sim{}R_S$ (counted from the initial moment of time $t=0$), the Bose-Einstein condensate of ``in'' excitations\footnote{Or, more precisely, quanta of states instantaneously diagonalizing the Hamiltonian of the scalar field $f$.} is formed. This condensate can be interpreted as the out-vacuum as it is the lowest energy state of the field at $t\to{}+\infty$. The form of the condensate is given by Eq. (\ref{Eq:QKAllenMottola}). For general conditions at $t=0$, it corresponds to generalized Mottola-Allen vacua of three-dimensional de Sitter space, the Bunch-Davies (``Euclidean'') vacuum being just one particular case. The word ``generalized'' used here means that the complex  quantity $\alpha$ parametrizing Mottola-Allen vacua can be an arbitrary function of in-momenta $k$.

We shall now discuss the implications of some of the observations listed above.
     
\subsection{Notes on the quantum kinetic equation}
\label{sec:QKnotes}

A quantum kinetic equation like (\ref{eq:kinetic_eq}) is an equation driving the evolution of occupation numbers $N_k(t)$ associated with a Hamiltonian $\hat{H}$ (roughly, $\hat{H}\sim\int{}dk\cdot{}\omega_k{}N_k$, where $\omega_k$ is the physical, observed frequency of excitations). As the Hamiltonian $\hat{H}$ can only be defined in QFTs on fixed  curved backgrounds up to time-dependent canonical/Bogolyubov transformations of the field \cite{Fulling:1979ac}, the occupation number $N_k$ by itself is not a physical, observable quantity. However, it is directly related to observables such as the expectation value of the stress-energy tensor of the field $f$ or its two-point correlation function. For example, in our setup the $(t,t)$-component of the stress-energy tensor is given at late times by
\begin{align}
\langle{}T_{tt}\rangle &= \frac{\sqrt{F(t)}}{2}\int d\omega_k \omega_k \bigl[(2N(t,\omega_k)+1)f_k(r)f_k^*(r)+\nonumber\\
&+X(t,\omega)f_k(r)f^*_{-k}(r)\bigr]+\nonumber\\
&+\frac{1}{2\sqrt{F(t)}}\int d\omega_k \omega_k \bigl[(2N(t,\omega_k)+1)f_k(r)f_k^*(r)-\nonumber\\
&-X(t,\omega)f_k(r)f^*_{-k}(r)\bigr],
\label{Eq:EMTVEV}
\end{align}
while the two-point function of the scalar field $f$ at equal (late) times is
\begin{align}
\langle{}f(t,r)f(t,r')\rangle =&\nonumber\\
=\int{}\frac{d\omega_k}{2\omega_k\sqrt{F(t)}}&\bigl[(2N(t,\omega_k)+1)f_k(r)f_k^*(r')-&\nonumber\\
-&X_k(t,\omega_k)f_k(r)f_{-k}^*(r')\bigr],
\label{Eq:2ptVEV}
\end{align}
where $N$ and $X$ are the solutions of Eqs. (\ref{N_t}) and (\ref{X_t}). 

Although both expressions (\ref{Eq:EMTVEV}) and (\ref{Eq:2ptVEV}) are not regularized, it is often useful to keep UV divergent terms in them explicitly (for example, in order to see how UV divergences build up with time). To regularize expressions (\ref{Eq:EMTVEV}) and (\ref{Eq:2ptVEV}), one either has to introduce a physical UV cutoff or introduce the adiabatic basis of eigenfunctions $f_k$ \cite{SZ,Fulling:1979ac,MottolaQK}.
    
Quantum kinetic equations similar to (\ref{eq:kinetic_eq}) exist (and their structure is nearly universal) for many physical effects involving time-dependence of a background field.  (See, for example, \cite{Schmidt:1998vi}, where a similar quantum kinetic equation is derived for the process of Schwinger pair production in a strong electromagnetic field.) 
In particular, such an equation can be derived for an arbitrary quantum field theory in an arbitrary background spacetime.
If a theory does possess an out-vacuum/vacua, described by a quantum kinetic equation similar to (\ref{eq:kinetic_eq}), then the equation admits a time-independent solution(s). Such solutions are attractors in the sense that general solutions of (\ref{eq:kinetic_eq}) approach them as $t\to{}+\infty$. The physical meaning of such attractor solutions is that they encode the effects of in-vacuum polarization (i.e. the development of a new vacuum, different from the "in" state) driven by the time-dependent background field.  

In principle, since Eq. (\ref{eq:kinetic_eq}) completely describes the physics of particle production in a background spacetime, its solutions also contain information about particle production, the Hawking effect in our particular case.\footnote{As usual, the fundamental difference between vacuum polarization and particle production effects is that the former vanishes when the background field is turned off (the out-vacuum coincides with the usual Minkowski $O(d-1,1)$-invariant vacuum), while the latter remains.} The Hawking effect is encoded in time-dependent corrections to the time-independent asymptotic solution of (\ref{eq:kinetic_eq}). These corrections to the numbers $N_k(t)$ of instantaneous excitations vanish as $t\to{}+\infty$, but their contribution to the expectation value of the energy-momentum tensor remains finite. For example, as can be seen from (\ref{Eq:EMTVEV}), corrections to the time-independent spectrum $N(\omega)$, which scale with time like $\exp{}(-t/2R_S)$, lead to a finite energy flux at spatial infinity and describe the Hawking effect. This observation also makes it manifest that the Hawking effect is subdominant with respect to the effect of Bose-Einstein condensation leading to the formation of out-vacuum.
\subsection{Properties of the out-vacuum and observables}
\label{sec:VacNotes}

Why can the out-vacuum (\ref{Eq:QKAllenMottola}) be associated with a \emph{Bose-Einstein condensate} of instantaneous excitations? In the context of condensed matter physics, one usually says that the Bose-Einstein condensate is created if the spectrum of bosonic (quasi-)particles acquires the form
\begin{equation}
n(k) = n_0\delta{}(k) + \frac{1}{e^{\frac{k}{T}}-1},
\label{eq:BEC}
\end{equation}
that is, when a sharp peak is developed at zero frequency and the occupation number of the zero mode is extremely large. We instead have a smooth distribution which behaves as a power law (\ref{Eq:QKPowerLaw}) at large frequencies and remains finite (but large) at $\omega_k=0$.

However, in real physical systems the distribution (\ref{eq:BEC}) is almost never seen in its canonical form. If one starts from an arbitrary non-equilibrium initial distrubution $n(t,k)$ (or just cools a system down), a so-called ``precondensate" is formed first \cite{Kagan} within the characteristic (short) time scale of the order of the mean free time in the interacting Bose gas describing the quasi-particles of the system. The precondensate is characterized by large (but finite) occupation numbers $n(k)$ of quasi-particles in the IR part of the spectrum, but the distribution $n(k)$ is non-singular. Instead, the behavior of $n(k)$ is smooth at $k\ll{}T$. 

Another characteristic feature of the precondensate is a strong quantum coherence among the states of quasiparticles constituting it. However, the global quantum phase of the BEC is not yet formed, and the time of its formation is much larger than the mean free time and actually depends on the system size \cite{Kagan}. To conclude, for a physical system with infinite $3$-volume the quantum state describing a precondensate is as close to the actual BEC as one can get even as $t\to+\infty$.  

We would like now to discuss thermal properties of the ``out'' vacua. The Bogolyubov coefficients between the ``in'' (Minkowski) vacuum and the Bunch-Davies vacuum (\ref{Eq:QKBunchDavies}) were calculated in  Sec.\ref{sec:CGHSBogolyubov}, and it was found that the corresponding spectrum of Hawking radiation is thermal with the correct temperature. Let us also calculate the detector response function for a fiducial observer located at $r\gg{}R_S$. For a detector interacting with the field according to
\begin{equation}
\int{}dt{}m(t)f(t,r(t))
\label{Eq:MonopoleMoment}
\end{equation}
(where $r=r(t)$ is the trajectory of the detector, and for us is a constant), 
one has \cite{Birrell:1982ix,Bousso:2001mw} for the probability of transition between the states $|E_i\rangle$ and $|E_j\rangle$ of the detector
\begin{equation}
\dot{P}_{ij}=|m_{ij}|^2\int^{+\infty}_{-\infty}dte^{-i\Delta{}Et}G(t,r(t);0,r(0)),
\label{Eq:ResponseFunction}
\end{equation}
where $G(x,y)$ is the Wightman function of the field $f(t,r)$ and $m_{ij}=\langle{}E_i|m(0)|E_j\rangle$. For the Bunch-Davies vacuum, the Wightman function $G$ is given by\footnote{It is worth emphasizing again that this $G$ differs from the Wightman function of the scalar field $f(t,r)$ in $dS_3$, because, for a given mode with momentum $k$, the phase volume spanned by the mode is different in our collapse problem and in $dS_3$. This is why we must compute the Wightman function explicitly instead of simply borrowing the known result for $dS_3$.}
\begin{equation}
G = \int_{R_S^{-1}}^{+\infty}{}dk{}\frac{\pi\eta\eta'}{4}H^{(2)\star}_1(k\eta)H^{(2)}_1(k\eta'),
\label{Eq:GWightman}
\end{equation}
where $\eta=-2R_S$, $\eta'=-2R_S\exp(-t/2R_S)$ are conformal times. Note that the lower limit of integration is taken to be $R_S^{-1}$, as the modes with lower momenta are out of equilibrium and are not in the Bunch-Davies vacuum state. The resulting expression (\ref{Eq:GWightman}) is periodic in imaginary time $t$ with the period $4\pi{}R_S$. Correspondingly, the detailed balance equation for the detector \cite{Bousso:2001mw} is
\begin{equation}
\frac{\dot{P}_{ij}}{\dot{P}_{ji}}=\exp{}(-4\pi{}R_S|E_i-E_j|)\equiv\exp{}(-4\pi{}R_S\Delta{}E),
\label{Eq:DetailedBalanceBD}
\end{equation}
and we conclude that the detector response is thermal with $T=T_H=\frac{1}{4\pi{}R_S}$.

This is no longer so when the out-vacuum of the field $f(t,r)$ is the generalized Mottola-Allen vacuum (\ref{Eq:QKAllenMottola}). Even if $\alpha_k=\alpha={\rm Const.}$, the detector response is not thermal \cite{Bousso:2001mw,Kaloper:2002cs}. Indeed, the Wightman function of the field $f$ is now given by
\begin{align}
G_\alpha(x,x')=\frac{1}{1-e^{\alpha+\alpha^*}}(G_E(x,x')+e^{\alpha+\alpha^*}G_E(x',x)+\nonumber\\
+e^{\alpha^*}G_E(x,x'_A)+e^\alpha{}G_E(x_A,x')),
\label{Eq:GAlphaWightman}
\end{align}
where $G_E$ is the Bunch-Davies Wightman function (\ref{Eq:GWightman}) and the index $A$ means that one takes a point on $dS_3$ antipodal to $x$. Correspondingly, the detailed balance equation is
\begin{equation}
\frac{\dot{P}_{ij}}{\dot{P}_{ji}}=e^{-4\pi{}R_S\Delta{}E}\left|\frac{1+e^{\alpha+2\pi{}R_S\Delta{}E}}{1+e^{\alpha-2\pi{}R_S\Delta{}E}}\right|^2,
\label{Eq:DetailedBalanceAM}
\end{equation}
i.e., the detector does not equilibrate, and the Hawking flux associated with the vacuum (\ref{Eq:QKAllenMottola}) is not thermal. This observation has   important implications for the information loss paradox as we discuss in Section \ref{sec:IL} below.

Given (\ref{Eq:DetailedBalanceAM}), the probability to excite states with $E_i\gg{}R_S^{-1}$ is finite for generalized Mottola-Allen vacua \cite{Kaloper:2002cs}, which means that the UV tail of the distribution of Hawking radiation is much heavier for such vacua than for the Bunch-Davies vacuum. \footnote{
It is possible to see immediately, without explicitly solving the quantum kinetic equation, that the spectrum $N(\omega_k)$ should behave as a power law at large frequencies. The equation for the amplitude $A_k(t)$ of the mode, rewritten in terms of the variable $\tau$ defined in (\ref{Eta_Sigma}), has the form of the equation of motion for a harmonic oscillator with the variable frequency $\Omega_k(\tau)$ defined in (\ref{omega_sigma}):
\begin{equation}
A_k''+\Omega^2_k(\tau)A_k=0.
\end{equation}
The occupation number $N_k(\tau)$ of instantaneous excitations coincides with the adiabatic invariant associated with this oscillator. For large physical frequencies $\omega_k\gg{}R_S^{-1}$, when the adiabatic approximation holds very well, the change in the value of the adiabatic invariant $\Delta{}N_k$ (roughly, the number of particles produced) is determined by the singularities of $\Omega_k^2(\tau)$ (and $\Omega{}'(\tau)$) in the complex plane of $\tau$ \cite{LL1}. If $\Omega_k^2(\tau)$ does not have singularities on the real axis, $\Delta{}N_k$ is exponentially small since its value is given by the singularity closest to the real axis. However, in our case $\Omega_k^2(\tau)\sim{}k^2/\tau$, so that there is a simple pole present at $\tau=0$. This pole corresponds to the location of the event horizon. In this case, $\Delta{}N_k$ decays only as a power law at large $k$. It is interesting to note that if the horizon is not ultimately formed (for example, due to backreaction effects), then the out-vacuum is Minkowski, the angular frequency $\Omega_k(\tau)$ has no singularities on the real $\tau$ axis, and we return to the exponentially small answer for $\Delta{}N_k$, reproducing Hawking's answer already at this simple level.}

\subsection{Implications for the information loss paradox}
\label{sec:IL}

The emerging picture sheds some new light on the information loss paradox and proposed ways to resolve it. We claim that, for both two-dimensional CGHS collapse and $(3+1)$-dimensional Schwarzschild spherically-symmetric collapse, the time-dependent amplitudes $A_k(t)$ of the modes of the scalar field $f(t,r)$ change with time as if the modes were effectively propagating in a $dS_3$ spacetime. The Hubble parameter of this effective spacetime is related to the horizon scale in the collapsing geometry  
\begin{equation}
H=\frac{1}{2R_S}.
\label{Eq:HConcl}
\end{equation}
In the usual quasi-de Sitter geometry describing an inflating universe, the modes are constantly leaving the horizon. At the same time, the value of $H$ keeps falling until the slow roll condition
\begin{equation}
\frac{|\dot{H}|}{H^2}\ll{}1
\end{equation}
ceases to hold, and inflation stops. During the subsequent era of power law Hubble expansion inflationary modes reenter the horizon and become the seeds for primordial perturbations of the matter density $\delta{}\rho/\rho$. If the power law Hubble expansion continues {\it ad infinitum}, all inflationary modes eventually reenter the horizon, so that the information about the initial state for inflation becomes completely recovered at $t\to+\infty$. The information loss paradox therefore does not appear in this situation.

What happens in the effective $dS_3$ space associated with the dynamics of modes on the collapsing background once the backreaction effects are taken into account? As the collapsing shell/black hole Hawking evaporates, the horizon scale $R_S$ decreases. But then the associated $H$ scale (\ref{Eq:HConcl}) \emph{increases}! This setup looks like inflation driven by an inflaton field $\phi$ that rolls up its potential $V(\phi)\sim{}H^2=(2R_S)^{-2}$ towards the super-Planckian scales instead of rolling down towards the minimum of $V(\phi)$. Correspondingly, the modes will keep leaving the horizon even more effectively at late times and will never reenter it again.


Despite this difficulty, we feel that our picture might provide an effective resolution of the information loss paradox. To see how, let us discuss again the structure of the generalized Mottola-Allen vacua (\ref{Eq:QKAllenMottola}). The late time general asymptotic solution of the quantum kinetic equation (\ref{eq:kinetic_eq}) is given by (\ref{Eq:QKAllenMottola}). Setting vacuum-like initial conditions at $t={}0$, one can fix only the combination $|a_k|^2+|b_k|^2$, not the complex coefficients $a_k$ and $b_k$. In order to determine the values of $a_k$ and $b_k$ separately, that is to find which particular vacuum among the generalized Mottola-Allen vacua (\ref{Eq:QKAllenMottola}) is realized at $t\to{}+\infty$, one must fix the 2-point correlation function of the field $f(t,r)$ together with $N_k(t=0)$ at $t=0$. Such correlation functions do not have to respect Minkowski, $E(2)$ or any other symmetries at $t\to{}0$ as they are only determined by the initial conditions for the field.

Imagine now that we initially have a thin collapsing shell with radius $R(t=0)=R_0$. The shell is allowed to have wiggles/internal structure, with information about them being effectively encoded at $t=0$ in the modes of the field $f(t,r)$ with $k\ll{}R_0^{-1}$. Note that the state of the field will not be vaccum-like for such modes. Let us also assume that a correlated EPR pair of quantum particles is initially placed right behind the shell, say, at $R=R_0+\delta{}R$. Information about the pair and its quantum correlations is encoded in the two-point function of the field $f(t,x)$, namely, in the amplitudes and the quantum phases of the modes with $k\sim{}l^{-1}\ll{}R_0^{-1}$. 

As we argue, once the collapse has started, the condensate of ``in'' particles (i.e, the out-vacuum) builds up during the time interval of the order $\sim{}R_S$. Note that the shell is still very far from its own apparent horizon at this moment. The out-vacuum is the generalized Mottola-Allen vacuum (\ref{Eq:QKAllenMottola}) parametrized by the complex quantity $\alpha_k$ trivially related to the complex coefficents $a_k$ and $b_k$ in (\ref{Eq:ASol}) and therefore encoding the information about wiggles/internal structure of the collapsing shell as well as the coherent pair of quanta following the shell. 


As was explained in the previous Section, such a vacuum emits a \emph{non-thermal} flux of Hawking quanta, which can therefore carry away information about $\alpha_k$. The UV tail of the distribution of emitted quanta is heavy \cite{Kaloper:2002cs}, so we expect this vacuum to evaporate away in finite time. Since it takes an infinite amount of time for the collapse to finalize in the slicing that we use, all the information about the wiggles/internal structure/quantum correlations of the shell as well as quantum correlations of the pair following the shell will be released to spatial infinity \emph{before} the event horizon/black hole is formed. 

Finally, the only information which survives at $t\to{}+\infty$ without getting released to spatial infinity will be the information about the total mass of the shell (in the case of $(3+1)$-dimensional Schwarzschild collapse) or the value of the cosmological constant (in the case of $(1+1)$-dimensional CGHS collapse). The associated out-vacuum will have the Bunch-Davies form (\ref{Eq:QKBunchDavies}) leading to the thermal flux of Hawking radiation from the shell.

The AdS/CFT correspondence in a certain sense seems to confirm this scenario. According to it, the process of gravitational collapse on an AdS background corresponds to the process of thermalization in the non-equilibrium $N=4$ Super-Yang-Mills (SYM) quark-gluon plasma \cite{Chesler:2008hg,Chesler:2010bi,CaronHuot:2011dr,Chesler:2011ds}. The $N=4$ SYM theory is a conformal field theory (CFT), and CFTs are known to take a very long time to thermalize starting from a non-equilibrium state. (See for example \cite{Micha:2004bv}, where thermalization of a classical scalar CFT is studied.) Typically, when one starts from an arbitrary non-equilibrium state in a CFT and follows the kinetics of equilibration, the spectrum of CFT excitations quickly (within the mean free time  for the plasma of CFT excitations) approaches time-independent non-thermal asymptotics. Such asymptotics are characterized by power-law specta $k^{-\gamma}$ of bosonic excitations, where the exponents $\gamma$ are determined by collision integrals of the kinetic equation describing such thermalization and, correspondingly, interactions of the CFT. Such asymptotic solutions have a very long life time (much longer than the mean free time), although the CFT of course ultimately thermalizes. This phenomenon has been called ``prethermalization'' in cosmology and in the heavy-ion-collisions literature. 
\newline\indent
We suspect that the prethermalization phenomenon is exactly what is seen in \cite{Chesler:2008hg,Chesler:2010bi,CaronHuot:2011dr,Chesler:2011ds}. The theory quickly (during one mean free time) reaches dynamic equilibration. This regime is characterized by components of the stress-energy tensor reaching their equilibrium asymptotic values, but continued non-thermal behavior of correlation functions.

To conclude, the scenario presented above by itself does not probably constitute the resolution of the information loss paradox as many steps which we discuss should be significantly better quantified. In particular, our understanding of the decay of Mottola-Allen vacua (the decay rate, the spectrum of decay products, etc.) should advance well beyond \cite{Kaloper:2002cs}, the analysis similar to the one performed in the present paper should be reproduced for different slicings since the Schwarzschild coordinate system does not cover the whole BH spacetime\footnote{Especially, it is necessary to determine what an infalling observer experiences.}. 

Our discussion only involves collapsing shells made of the very same matter content as the most rapidly emitted modes of Hawking radiation (massless scalars). It would be instructive to understand what happens in the case of collapsing shells made of dust-like matter (for example, massive scalar with $m>R_S^{-1}$) and where exactly the information about their internal structure is encoded in Hawking radiation.\footnote{A preliminary analysis \cite{Greenwood:2010sx} shows that the structure of $M_{kk'}$ and $N_{kk'}$ kernels (see Section \ref{sec:CGHSmodes}) for massive scalars is not very different from the one for massless scalars (the case discussed in this paper). Therefore, we expect that the quantum kinetic equation for the number of instantaneous excitations will have the form analogous to (\ref{eq:kinetic_eq}), and our conclusions concerning the structure of the out-vacuum will be also similar. Then, the information about the initial state of the shell made of massive scalar particles will be encoded in the parameters of the generalized Mottola-Allen vacuum realized at $t\to+\infty$. The associated Hawking radiation will be strongly non-thermal, and the information about the initial state of the system will be released before the collapse is finalized.}

Ultimately, one must understand the behavior and origin of black-hole entropy using the formalism developed here. 
Nevertheless, we expect that this work is an important step towards the ultimate resolution of the celebrated information loss paradox.

\subsubsection*{Acknowledgements}

The authors would like to thank C. Kiefer, G. Kunstatter, E. Mottola, D. Stojkovic, A. Tolley, T. Vachaspati and G. Volovik for the discussions. This work is supported by the Department of Energy, through a grant to the particle astrophysics theory group at CWRU.   
Numerical calculations were carried out on the HPC at CWRU.

\appendix

\section{Massive shell collapsing on a primordial black hole}
\label{sec:PBH}

In this Appendix we first find the classical equations of motion for a massive shell collapsing onto a pre-existing primordial black hole (PBH). After which, we set up the equations for a scalar field coupled to the background of the collapsing shell around the PBH.

\subsection{Classical solution}

Here we take that the internal metric is that of a PHB, so that the metric is just described by the
Schwarzschild metric
\be
  ds_-^2=-\left(1-\frac{2GM_1}{r}\right)dT^2+\frac{dr^2}{1-2GM_1/r}+r^2d\Omega^2
  \label{met_int}
\ee
where $M_1$ is the mass of the PHB. The exterior metric is also given by the Schwarzschild metric,
however the mass here is now the total mass, i.e.~that of the PHB and the shell (with mass $M_2$):
\be
  ds_+^2=-\left(1-\frac{2GM_{tot}}{r}\right)dT^2+\frac{dr^2}{1-2GM_{tot}/r}+r^2d\Omega^2
  \label{met_ex}
\ee
where $M_{tot}=M_1+M_2$.

For a given interior metric with metric coefficient $J(r=R)$ and exterior metric with metric coefficient
$P(r=R)$, one can find the perpendicular components of the acceleration on either side of the shell.
Taking the ratio of the sum and difference of these once can find 
\be
  4\pi\sigma G(\alpha+\beta)=\frac{2(\alpha^2-\beta^2)}{R}+J'-P',
  \label{ratio}
\ee
where $\sigma$ is the energy density of the domain wall and
\be
  \alpha\equiv\sqrt{J+R_{\tau}^2}\hs \text{and} \hs \beta\equiv\sqrt{J+R_{\tau}^2}.
\ee
From (\ref{met_int}) and (\ref{met_ex}) we have that
\begin{align}
  J&=1-\frac{2GM_1}{R}\label{A},\\
  P&=1-\frac{2GM_{tot}}{R}=1-\frac{2G(M_1+M_2)}{R}\label{f}.
\end{align}
Substituting (\ref{A}) and (\ref{f}) into (\ref{ratio}) and solving for the mass of the shell gives:
\be
  M_2=\frac{2\pi\sigma R^2}{3}\left(\sqrt{J+R_{\tau}^2}+\sqrt{P+R_{\tau}^2}\right)
  \label{M_2 1}
\ee
or solving explicitly for the mass
\be
  M_2=\frac{4\pi\sigma R^2}{9}\left(3\sqrt{1-\frac{2GM_1}{R}+R_{\tau}^2}-2\pi\sigma GR\right).
  \label{M_2}
\ee
A mass of this form is actually a constant of motion, i.e.~$dM_2/d\tau=0$. Therefore, we can take $M_2=H$ to be the Hamiltonian of the system. Note that the Hamiltonian is in terms of the infalling observer time $\tau$, an observer who is riding on the shell as it is collapsing. What we want, for Cosmological purposes, is the Hamiltonian in
terms of the asymptotic observer time $t$. To get this, we must first find the Lagrangian of the system to
determine the action, then perform a coordinate transformation on the action, since the Hamiltonian is
not invariant under coordinate transformations, and then find the new Hamiltonian. To find the
coordinate transformation we use (\ref{met_ex}) and find that the relationship between $\tau$ and $t$ 
is
given by
\be
  \frac{dt}{d\tau}=\frac{1}{P}\sqrt{P+R_{\tau}^2}.
  \label{dtdtau}
\ee
The Hamiltonian as observed by the asymptotic observer is then given by:
\be
  H=\frac{4\pi\sigma R^2P^{3/2}}{9\sqrt{P^2-\dot{R}^2}}\left(3\sqrt{\frac{JP^2-(J-P)\dot{R}^2}{P^2-\dot{R}^2}}-2\pi\sigma GR\right).
  \label{full_H}
\ee
We are mostly interested in the near horizon limit, $R\to R_S=2GM_{tot}$, so that in this limit
$J\to const$ and $P\to0$. Taking the near horizon limit of (\ref{full_H}) we then have
\be
  H=\frac{4\pi\mu R^2f^{3/2}}{9\sqrt{f^2-\dot{R}^2}}
  \label{H}
\ee
where
\begin{align}
  \mu&\equiv\sigma\left(3\sqrt{J(R=R_S)}-2\pi\sigma GR_S\right)\nonumber\\
    &=\sigma\left(3\sqrt{1-\frac{2GM_1}{R_S}}-2\pi\sigma GR_S\right)
    \label{mu}
\end{align}

To find the radius of the shell as a function of asymptotic observer time $t$, we need to solve (\ref{H}) for $\dot{R}$, which gives
\be
  \dot{R}=\pm P\sqrt{1-\frac{R^4}{h^2}P}
  \label{dot R full}
\ee
where $h\equiv 9H/(4\pi\mu)$. In the near horizon limit we can then write
\be
  \dot{R}=\pm P\left(1-\frac{1}{2}\frac{R^4}{h^2}P\right).
\ee
Therefore the dynamics of shell in the near horizon regime can be obtained by solving $\dot{R}=-P$,
where we took the minus sign since the shell is collapsing, which is identical to that in
\cite{Vachaspati:2006ki}.

\subsection{Quantum radiation of instantaneous excitations}

Here we follow the procedure found in \cite{Vachaspati:2006ki}, where we split the action in to two
parts, one for the interior and the other for the exterior, and find the dominant pieces. For the interior
metric we have
\begin{align}
  S_{-}=2\pi\int dt\int_0^{R_S}dr\,r^2\Big{[}&-\frac{1}{\dot{T}}\frac{(\partial_t\Phi)^2}{1-2GM_1/r}\nonumber\\
    &+\dot{T}\left(1-\frac{2GM_1}{r}\right)(\partial_r\Phi)^2\Big{]}
\end{align}
and the exterior we have
\begin{align}
  S_{+}=2\pi\int dt\int_{R_S}^{\infty}dr\,r^2\Big{[}&-\frac{(\partial_t\Phi)^2}{1-2GM_{tot}/r}\nonumber\\
    &+\left(1-\frac{2GM_{tot}}{r}\right)(\partial_r\Phi)^2\Big{]}.
\end{align}
To move on, we must determine $\dot{T}=(dT/d\tau)(d\tau/dt)$. From the structure of the metric, the
relationship between $\tau$ and $T$ is analogous to that of $\tau$ and $t$ (\ref{dtdtau}), which is
given by
\be
  \frac{dT}{d\tau}=\frac{1}{J}\sqrt{J+R_{\tau}^2}.
  \label{dTdtau}
\ee
Therefore using (\ref{dtdtau}), (\ref{dTdtau}) and (\ref{dot R full}) we find that
\be
  \dot{T}=\frac{1}{J}\sqrt{JP-\frac{J-P}{P}\dot{R}^2}=\frac{P}{J}\sqrt{1+(J-P)\frac{R^4}{h^2}}
\ee
so that in the near horizon limit $\dot{T}\to P/J\sim P$, since in this limit $J$ is finite. Comparing terms and keeping only the dominant contributions, we can then write the total action as
\begin{align}
  S\approx2\pi\int dt\Big{[}&-\frac{1}{P}\int_0^{R_S}dr\,r^2\frac{(\partial_t\Phi)^2}{1-2GM_1/r}\nonumber\\
    &+\int_{R_S}^{\infty}dr\,r^2\left(1-\frac{R_S}{r}\right)(\partial_r\Phi)^2\Big{]}
\end{align}

As in \cite{Vachaspati:2006ki} one can expand $\Phi$ in modes and define matrices $\tilde {\bf M}$
and $\tilde {\bf N}$ that are independent of $R(t)$, here defined by
\begin{align}
  \tilde {\bf M}_{kk'}=4\pi\int_0^{R_S}dr\,r^2\frac{f_k(r)f_{k'}(r)}{1-2GM_1/r},\\
  \tilde {\bf N}_{kk'}=4\pi\int_{R_S}^{\infty}dr\,r^2\left(1-\frac{R_S}{r}\right)f'_k(r)f'_{k'}(r).
\end{align}
Using the standard quantization procedure and performing a principle axis transformation, we find
that for a single eigenmode, the Schr\"odinger equation takes the form
\be
  \left[-\frac{1}{2\tilde m}\frac{\partial^2}{\partial b^2}+\frac{\tilde m}{2}\tilde{\omega}^2(\eta)\right]\psi(b,\eta)=i\frac{\partial\psi(b,\eta)}{\partial\eta}
\ee
where $\tilde m$ denotes the eigenvalue of $\tilde {\bf M}$, $b$ is the eigenmode,
\be
  \tilde \omega^2(\eta)=\frac{\tilde K}{\tilde m}\frac{1}{1-R_S/R}\equiv\frac{\tilde \omega_0^2}{1-R_S/R}
\ee
where $\tilde K$ denotes the eigenvalue of $\tilde {\bf N}$ and
\be
  \eta=\int_0^tdt'\left(1-\frac{R_S}{R}\right).
\ee
This has identical structure to that found for the massive shell collapsing upon itself.

\section{Collapse of two shells}
\label{sec:2Shells}

In this Appendix we will again first find the classical equations of motion for two massive shells
collapsing. After which, we set up the equations for a scalar field coupled to the background of the
collapsing shells.

\subsection{Classical Solution}

\begin{figure}[t]
\includegraphics[width=0.55\textwidth,height=0.3\textheight]{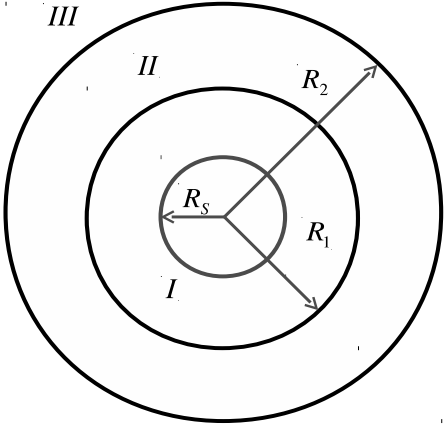}
\caption{Schematic of the collapse of two shells. The interior of shell 1, with mass $M_1$ at radius $R_1$ is taken to be Minkowski, the region between the two shells is taken to be Schwarzschild and the exterior of shell 2, with mass $M_2$ at radius $R_2$ is taken to be Schwarzschild.}
\label{2Shells}
\end{figure}

Consider two concentric shells collapsing under gravitation, where the inner shell is located at a radius $R_1$ with mass $M_1$ and the outer shell is located at radius $R_2$ (where $R_2>R_1$) with mass $M_2$. Thus we have three regions denoted by $I$, $II$, and $III$, see Figure \ref{2Shells}. Due to Birkhoff's theorem, the metric in region $I$ is taken to be flat Minkowski
\be
  ds_I^2=-dT^2+dr^2+r^2d\Omega^2,
\ee
the metric in region $II$ is taken to be Schwarzschild
\be
  ds_{II}^2=-\left(1-\frac{2GM_1}{r}\right)d\tilde t^2+\frac{dr^2}{1-2GM_1/r}+r^2+d\Omega^2,
\ee
and the metric in region $III$ is taken to also be Schwarzschild
\be
  ds_{III}^2=-\left(1-\frac{2GM_{tot}}{r}\right)dt^2+\frac{dr^2}{1-2GM_{tot}/r}+r^2+d\Omega^2,
\ee
where $M_{tot}=M_1+M_2$. Due to the structure of the spacetime, each of the shells satisfy the Gauss-Codazzi equation separately. Thus considering the shells individually we find that mass $M_1$, which is a constant of motion, is given by
\be
  M_1=4\pi\tilde\sigma R_1^2\left(\sqrt{1+\left(\frac{dR_1}{d\tilde \tau}\right)^2}-2\pi\tilde\sigma GR_1\right)
  \label{2shell M1}
\ee
where $\tilde \tau$ is the proper time of an observer riding on the shell. This result is not unexpected
since it is just the case considered by Ipser and Sikivie in \cite{Ipser:1983db}. Here the internal time
$T$ is related to the proper time $\tilde \tau$ by
\be
  \frac{dT}{d\tilde \tau}=\sqrt{1+\left(\frac{dR_1}{d\tilde \tau}\right)^2}
  \label{dTdttau}
\ee
and the ``external" time $\tilde t$ is related to $\tilde \tau$ by
\be
  \frac{d\tilde t}{d\tilde \tau}=\frac{1}{B_1}\sqrt{B_1+\left(\frac{dR_1}{d\tilde \tau}\right)^2}
  \label{dttdttau}
\ee
where
\be
  B_1\equiv1-\frac{2GM_1}{R_1}.
  \label{B1}
\ee
We find that mass $M_2$, which is also a constant of motion, is given by
\be
  M_2=\frac{4\pi\sigma R_2^2}{9}\left(3\sqrt{B_2+\left(\frac{dR_2}{d\tau}\right)^2}-2\pi\sigma GR_1\right)
\ee
where $\tau$ is the proper time of an observer riding on the shell and
\be
  B_2\equiv1-\frac{2GM_1}{R_2}.
  \label{B2}
\ee
Again this result is not unexpected since this is the scenario examined in Appendix \ref{sec:PBH},
see (\ref{M_2}). The ``internal" time $\tilde t$ and external time $t$ are related to the propertime $\tau$
by
\begin{align}
  \frac{d\tilde t}{d\tau}=\frac{1}{B_2}\sqrt{B_2+\left(\frac{dR_2}{d\tau}\right)^2} \label{dttdtau}\\
  \frac{dt}{d\tau}=\frac{1}{B}\sqrt{B+\left(\frac{dR_2}{d\tau}\right)^2} \label{dtdtau1}
\end{align}
where
\be
  B\equiv1-\frac{2GM_{tot}}{R_2}.
\ee

Since the masses are separately conserved, we can take them as the Hamiltonians of the shells. We
are interested in the collapse as observed by the asymptotic observer, hence we must find the
equation of motion of the two shells as a function of time $t$. We will first consider the outer shell,
$M_2$.

It is easy to see that there is no difference in the evaluation of the equation of motion as a function of
time $t$ in the two shell case than that of the shell collapsing onto the PBH. Therefore we find that in
the near horizon limit, where here that corresponds to $B\to0$, the Hamiltonian corresponds to
\be
  H_2=\frac{4\pi\mu R_2^2B^{3/2}}{9\sqrt{B^2-\dot{R}_2^2}}
\ee
which is just (\ref{H}) with the appropriate substitutions and $\mu$ is defined in (\ref{mu}). To leading
order in $B$, the equation of motion for the position of the shell is then given by
\be
  \dot{R}_2\approx-B
  \label{dot R2}
\ee

Now we consider the equation of motion for the inner shell of mass $M_1$. The Hamiltonian is a
function of $\tilde \tau$, where we want as a function of $t$. From (\ref{dttdttau}) we see that
$\tilde \tau$ is a related to $\tilde t$, while from (\ref{dttdtau}) and (\ref{dtdtau1}) $\tilde t$ is related to
$t$ via $\tau$. Therefore we can perform the coordinate transformation
\be
  \frac{d\tilde \tau}{dt}=\frac{d\tilde \tau}{d\tilde t}\frac{d\tilde t}{d\tau}\frac{d\tau}{dt}
  \label{dttaudt}
\ee
to transform (\ref{2shell M1}) from a function of $\tilde \tau$ to a function of $t$. By first finding the
corresponding Lagrangian and determining the action, then performing the coordinate transformation
given in (\ref{dttaudt}), we find that the Hamiltonian as a function of the asymptotic observer time $t$
is given by
\begin{align}
  H_1=&\frac{4\pi\tilde \sigma R_1^2B_1^{3/2}}{B_2\sqrt{B}}\frac{B_2B^2-(B_2-B)\dot{R}_2^2}{\sqrt{B^2B_1^2B_2-BB_2^2\dot{R}_1^2-(B_2-B)\dot{R}_2^2}}\nonumber\\
  &\times\Big{(}\sqrt{\frac{B^2B_1^2B_2-(1-B_1)BB_2^2\dot{R}_1^2-B_1^2(B_2-B)\dot{R}_2^2}{B^2B_1^2B_2-BB_2^2\dot{R}_1^2-B_1^2(B_2-B)\dot{R}_2^2}}\nonumber\\
  &-2\pi\tilde \sigma GR_1\Big{)}.
  \label{H1t1}
\end{align}
We will again be concerned with the near horizon limit, where here that corresponds to $B_1\to0$,
which is inside of the actual Schwarzschild radius $R_S=2GM_{tot}$. However, until the
Schwarzschild radius is formed, the asymptotic observer will be able to see this region. Taking this
limit, we see that for the squareroot in the parenthesis, the numerator and denominator are the same,
so (\ref{H1t1}) can be written as
\begin{align}
  H_1=&\frac{4\pi\tilde \sigma R_1^2B_1^{3/2}}{B_2\sqrt{B}}\frac{B_2B^2-(B_2-B)\dot{R}_2^2}{\sqrt{B^2B_1^2B_2-BB_2^2\dot{R}_1^2-(B_2-B)\dot{R}_2^2}}\nonumber\\
  &\times\left(1-2\pi\tilde \sigma GR_1\right)\nonumber\\
  \equiv&\frac{4\pi\tilde \mu R_1^2B_1^{3/2}}{B_2\sqrt{B}}\frac{B_2B^2-(B_2-B)\dot{R}_2^2}{\sqrt{B^2B_1^2B_2-BB_2^2\dot{R}_1^2-(B_2-B)\dot{R}_2^2}}
  \label{H1t2}
\end{align}
where
\be
  \tilde \mu\equiv\tilde\sigma\left(1-2\pi\tilde\sigma GR_S^1\right)
\ee
and $R_S^1=2GM_1$. Here we see that (\ref{H1t2}) depends explicitly on $\dot{R}_2$. To simplify
things further we will make the assumption that $R_1$ and $R_2$ don't start that far away from each
other so that we can treat that $R_2$ is simultaneously in the near horizon limit. Thus we take that
$\dot{R}_2$ satisfies (\ref{dot R2}) that that (\ref{H1t2}) can be written as
\be
  H_1\approx\frac{4\pi\tilde\mu R_1^2B^{5/2}B_1^{3/2}}{B_2\sqrt{B^2B_1^2-B_2^2\dot{R}_1^2}}.
\ee
Solving for $\dot{R}_1$ yields
\be
  \dot{R}_1=-B_1\frac{B}{B_2}\sqrt{1-\frac{B_1R_1^4B^3}{B_2^2h_1^2}}
  \label{dot R11}
\ee
where $h_1\equiv H_1/(4\pi\tilde\mu)$. Since $B_1\to0$ we can Taylor expand and see that the
dynamics of $R_1$ are given by
\be
  \dot{R}_1\approx-B_1\frac{B}{B_2}.
  \label{dot R1}
\ee

Solving (\ref{dot R1}) to lowest order in $R_0^{(1)}-R_S^{(1)}$ we find, using the explicit time dependence of $B$,
\be
  R_1\approx R_S^{(1)}+(R_0^{(1)}-R_S^{(1)})\exp\left[\frac{R_S(R_0^{(2)}-R_S)}{R_S^{(1)}R_S^{(2)}}(e^{-t/R_S}-1)\right]
  \label{R_1 Sol}
\ee
where the superscripts in parenthesis mean the initial position of the corresponding mass. Hence we
see that as $t\to\infty$, $R_1\to R_S^{(1)}+const$, so that $B_1$ never quite reaches zero, however it
can be arbitrarily close.
\subsection{Quantum radiation of instantaneous excitations }
In this case, the action for the scalar field coupled to the collapse of the two shells is then split into
three pieces. We again must find the dominate parts. The individual actions then takes the form
\bd
  S_I=2\pi\int dt\int_0^{R_1}dr\,r^2\Big{[}-\frac{1}{\dot{T}}(\partial_t\Phi)^2+\dot{T}(\partial_r\Phi)^2\Big{]}
\ed
\begin{align*}
  S_{II}=2\pi\int dt\int_{R_1}^{R_2}dr\,r^2\Big{[}&-\frac{1}{\dot{\tilde t}}\frac{(\partial_t\Phi)^2}{1-2GM_1/r}\\
       &+\dot{\tilde t}\left(1-\frac{2GM_1}{r}\right)(\partial_r\Phi)^2\Big{]}
\end{align*}
and
\begin{align}
  S_{III}=2\pi\int dt\int_{R_2}^{\infty}dr\,r^2\Big{[}&-\frac{(\partial_t\Phi)^2}{1-2GM_{tot}/r}\\
       &+\left(1-\frac{2GM_{tot}}{r}\right)(\partial_r\Phi)^2\Big{]}
\end{align}
where we transformed to the asymptotic observer time $t$. Using (\ref{dot R2}) and (\ref{dot R2}) we
find that $\dot T\approx BB_1/B_2$ and $\dot{\tilde t}\approx B/B_2$. Therefore we can write
\be
  S_I=2\pi\int dt\int_0^{R_1}dr\,r^2\Big{[}-\frac{B_2}{BB_1}(\partial_t\Phi)^2+\frac{BB_1}{B_2}(\partial_r\Phi)^2\Big{]}
  \label{SI}
\ee
\begin{align}
  S_{II}=2\pi\int dt\int_{R_1}^{R_2}dr\,r^2\Big{[}&-\frac{B_2}{B}\frac{(\partial_t\Phi)^2}{1-2GM_1/r}\nonumber\\
       &+\frac{B}{B_2}\left(1-\frac{2GM_1}{r}\right)(\partial_r\Phi)^2\Big{]}
   \label{SII}
\end{align}
and
\begin{align}
  S_{III}=2\pi\int dt\int_{R_2}^{\infty}dr\,r^2\Big{[}&-\frac{(\partial_t\Phi)^2}{1-2GM_{tot}/r}\nonumber\\
       &+\left(1-\frac{2GM_{tot}}{r}\right)(\partial_r\Phi)^2\Big{]}
   \label{SIII}
\end{align}
Here we will make some comments on (\ref{SI}) - (\ref{SIII}). First, note that in the limit $M_1\to0$ the
upper limit of integrate of (\ref{SI}) goes to $R_2$, while the lower limit of (\ref{SII}) also goes to $R_2$.
Thus (\ref{SI}) reduces to (44) in \cite{Vachaspati:2006ki}, while (\ref{SII}) goes to zero. Thus we recover
the scenario investigated in \cite{Vachaspati:2006ki}. Second, note that in the limit $M_2\to0$,
$B_1=B_2=B$, so that again (\ref{SI}) reduces to (44) in \cite{Vachaspati:2006ki}, while again (\ref{SII})
goes to zero and (\ref{SIII}) reduces to (45) in \cite{Vachaspati:2006ki}. Again we recover the scenario
investigated in \cite{Vachaspati:2006ki}.

We are now in a position to determine the dominate parts of the action for the scalar field. First we
consider the gradient term. We see that in the $R_2\to R_S$ region both the gradient terms in $S_I$
and $S_{II}$ go to zero since $B\to0$, however, the gradient term in $S_{III}$ goes to a finite number,
thus we keep the gradient term from (\ref{SIII}). For the kinetic term, more care is needed to determine
which term dominates. For $S_{III}$ we see that the kinetic term is logarithmically divergent, while both
$S_I$ and $S_{II}$ are $(R_2-R_S)^{-1}$ divergent, due to the factor of $B$. However, we also see
that the kinetic term in $S_I$ goes as $(R_1-R_S^{(1)})^{-1}$, while $S_{II}$ is logarithmically
dependent. Hence, since $R_1$ can get arbitrarily close to $R_S^{(1)}$ we can conclude that the
kinetic term in $S_I$ is the dominant term. Hence the total action can be written as
\begin{align}
  S\sim2\pi\int dt\Big{[}&-\frac{B_2}{BB_1}\int_0^{R_S^{(1)}}dr\,r^2(\partial_t\Phi)^2\nonumber\\
         &+\int_{R_S}^{\infty}dr\,r^2\left(1-\frac{2GM_{tot}}{r}\right)(\partial_r\Phi)^2\Big{]}
\end{align}

As in \cite{Vachaspati:2006ki} one can expand $\Phi$ in modes and define matrices $\tilde {\bf A}$ and
$\tilde {\bf C}$ that are independent of $R_1(t)$ and $R_2(t)$, here defined by
\begin{align}
  \tilde {\bf A}_{kk'}=4\pi\int_0^{R_S^{(1)}}dr\,r^2f_k(r)f_{k'}(r),\\
  \tilde {\bf C}_{kk'}=4\pi\int_{R_S}^{\infty}dr\,r^2\left(1-\frac{R_S}{r}\right)f'_k(r)f'_{k'}(r).
\end{align}
Once again using the standard quantization procedure and performing a principle axis transformation,
we find that for a single eigenmode, the Schr\"odinger equation takes the form
\be
  \left[-\frac{1}{2\bar m}\frac{\partial^2}{\partial b^2}+\frac{\bar m}{2}\bar{\omega}^2(\eta)\right]\psi(b,\eta)=i\frac{\partial\psi(b,\eta)}{\partial\eta}
\ee
where $\bar m$ denotes the eigenvalue of $\tilde {\bf A}$, $b$ is the eigenmode,
\be
  \bar \omega^2(\eta)=\frac{\bar K}{\bar m}\frac{B_2}{BB_1}\equiv\frac{\bar \omega_0^2B_2}{BB_1}
\ee
where $\bar K$ denotes the eigenvalue of $\tilde {\bf C}$ and
\be
  \eta=\int_0^tdt'\frac{BB_1}{B_2}.
\ee
This has identical structure to that found for a single massive shell collapsing upon itself and a massive shell collapsing onto a PBH.

\section{Notes on the breakdown of the WKB expansion for the Wheeler-de Witt equation}
\label{sec:WKBnotes}

One implication of our results to which is worth paying special attention is related to a famous issue arising in first quantization of a scalar field on a collapsing spacetime \cite{Kiefer:1990pt,Demers:1995tr,Greenwood:2010sx}. When one performs such quantization, one looks for a wave functional $\Psi$ being a solution of the Wheeler-de Witt equation $\hat{H}_{\rm WdW}\Psi=0$ for the scalar field (+ moving shell + gravitational degrees of freedom) with vacuum-like initial conditions. The square of this wave functional $|\Psi|^2$ then provides the probability density to find the shell and emitted Hawking radiation in a given state. 

In $(3+1)$-dimensions one can look for the solution of the Wheeler-de Witt equation in the form $\Psi = e^{iS}$ with the ``action'' expanded as
\be
S = G^{-1}S_0 + S_1 + GS_2 + \cdots, \label{eq:SWKB}
\ee
where $G$ is the gravitational constant. The expansion (\ref{eq:SWKB}) is nothing but a WKB expansion for the wave functional. The zeroth order term denoted $S_0$ is the solution of the Einstein-Hamilton-Jacobi equation for the background spacetime, the $S_1$ term describes effects of quantum theory of the scalar field $f$ on the fixed background spacetime (it therefore includes the Hawking effect) and the wave functional $e^{iS_1}$ satisfies the functional Schr\"{o}dinger equation used in the Sec. \ref{sec:CGHSQuantumKineticOUT}, the $S_2$ term describes the backreaction of the background  to the Hawking radiation  in the leading approximation in $1/M_P$, etc. In the $(1+1)$-dimensional CGHS case, a similar WKB expansion of the wave functional exists: it is either the expansion in terms of the dimensionless gravitational constant or, equivalently, the expansion in terms of the parameter $\kappa = \frac{N-24}{6}$, where $N$ is the (large) number of matter fields $f$ \cite{Demers:1995tr}. 

The expansion (\ref{eq:SWKB}) has been proved to break down in the near horizon regime \cite{Kiefer:1990pt,Demers:1995tr}. 
This effect seems to be puzzling as it cannot be seen in the second-quantized version of the theory. Indeed, the standard intuition suggests that, once the shell is close to its apparent horizon, the applicability of the geometric-optics approximation for the modes of $f$ improves, and the phases of the modes pile up. Correspondingly, one expects that the phase of the wave functional $\Psi$ oscillates rapidly, i.e, the WKB approximation should be better, not worse.

It is possible to show using the formalism of the quantum kinetic equation that the WKB breakdown effect noticed in \cite{Kiefer:1990pt} can be mostly attributed to the Bose-Einstein condensation of ``in'' excitations and the formation of the out-vacuum. As we have discussed, the number of instantaneous excitations at late times $t\to{}+\infty$ is given by the solution of the quantum kinetic equation (\ref{eq:kinetic_eq}) and can be represented in the form of an expansion in powers of $\exp{}(-t/2R_S)$:
\begin{equation}
N(t,\omega_k{})=N(\omega_k{})+n(\omega_k{})e^{-t/2R_S}+O(e^{-t/R_S}).
\label{Eq:Nexpansion}
\end{equation}
In this expansion, the first term contributes to the polarization of ``in"-vacuum (i.e, formation of the out-vacuum) and, correspondingly, to the redefinition of $S_0$ in (\ref{eq:SWKB})). The second term proportional to $n(\omega_k)$ describes the Hawking flux (since it corresponds to the finite energy flux at spatial infinity, see (\ref{Eq:EMTVEV}) and subsequent discussion) and therefore contributes to $S_1$ in (\ref{eq:SWKB}). Once the vacuum is properly redefined, the zeroth order contribution $\sim{}S_0$ in (\ref{eq:SWKB}) becomes large compared to the first order contribution $\sim{}S_1$, because the second term in (\ref{Eq:Nexpansion}) is subdominant with respect to the first term as $t\to+\infty$.

We finally note that the WKB breakdown effect cannot be entirely attributed to the vacuum polarization, as in the slicing we use one expects the quantum gravitational effects of ``horizon trembling'' to become important in the near horizon regime \cite{Trembling1,Trembling2,Trembling3}.

\end{document}